\documentclass[sigconf]{acmart}
\usepackage{multirow}
\usepackage{stfloats}
\usepackage{subcaption}
\usepackage{array}
\newcolumntype{M}[1]{>{\centering\arraybackslash}m{#1}}

\AtBeginDocument{%
  \providecommand\BibTeX{{%
    \normalfont B\kern-0.5em{\scshape i\kern-0.25em b}\kern-0.8em\TeX}}}

\copyrightyear{2023}
\acmYear{2023}
\setcopyright{rightsretained}
\acmConference[CHI '23]{Proceedings of the 2023 CHI Conference on Human Factors in Computing Systems}{April 23--28, 2023}{Hamburg, Germany}
\acmBooktitle{Proceedings of the 2023 CHI Conference on Human Factors in Computing Systems (CHI '23), April 23--28, 2023, Hamburg, Germany}
\acmDOI{10.1145/3544548.3581425}
\acmISBN{978-1-4503-9421-5/23/04}
\begin{document}

%\title[Counterbalancing Visual Privacy Awareness...]{Counterbalancing Visual Privacy Awareness and Activities of Daily Living (ADLs) on Low-resolution Imagers}

%Optimizing Visual Privacy Preserving Machine Recognition on Low-resolution Image Sensors

\title[Modeling the Trade-off of Privacy ...]{Modeling the Trade-off of Privacy Preservation and Activity Recognition on Low-Resolution Images}

%\title[Exploring the Effect of Image Resolution ...]{Exploring the Effect of Image Resolution on Privacy-Preserving Machine Recognition}%of Activities of Daily Living (ADLs)}

% \title[Exploring Low Resolution Image Privacy Preserving  ...]{Exploring the Effect of Image Resolution on Visual Privacy Preserving Machine  Recognition of Activities of Daily Living (ADLs)}

%\title[Finding the Sweet Spot]{Understanding the Effect of Image Resolution on Visual Privacy Preserving Activities of Daily Living (ADLs) Recognition}

%Finding the Sweet Spot: Modeling the Effect of Image Resolution for Visual Privacy Preserving Machine Recognition in Daily Living Environments


\author{Yuntao Wang}
\authornote{The authors contribute equally to this paper.}
\email{yuntaowang@tsinghua.edu.cn}
\affiliation{
  \institution{Key Laboratory of Pervasive Computing, Ministry of Education, \\ Department of Computer Science and Technology, \\ Tsinghua University}
  \city{Beijing}
  \country{China}}

\author{Zirui Cheng}\authornotemark[1]
\email{chengzr19@mails.tsinghua.edu.cn}
\affiliation{
  \institution{Department of Computer Science and Technology, \\ Tsinghua University}
  \city{Beijing}
  \country{China}}

\author{Xin Yi}\authornote{denotes as the corresponding author.}
\email{yixin@tsinghua.edu.cn}
\affiliation{
  \institution{Institute for Network Sciences and Cyberspace, \\ Tsinghua University}
  \city{Beijing}
  \country{China}}
\affiliation{
    \institution{Zhongguancun Laboratory}
    \city{Beijing}
    \country{China}
}

\author{Yan Kong}
\email{ky21@mails.tsinghua.edu.cn}
\affiliation{
  \institution{Institute for Network Sciences and Cyberspace, \\ Tsinghua University}
  \city{Beijing}
  \country{China}}

\author{Xueyang Wang}
\email{wang-xy22@mails.tsinghua.edu.cn}
\affiliation{
  \institution{Institute for Network Sciences and Cyberspace,\\ Tsinghua University}
  \city{Beijing}
  \country{China}}

\author{Xuhai Xu}
\email{xuhaixu@cs.washington.edu}
\affiliation{
  \institution{Information School, \\ University of Washington}
  \city{Seattle, WA}
  \country{USA}}

\author{Yukang Yan}
\email{yanyukanglwy@gmail.com}
\affiliation{
  \institution{Department of Computer Science and Technology, \\ Tsinghua University}
  \city{Beijing}
  \country{China}}

\author{Chun Yu}
\email{chunyu@mail.tsinghua.edu.cn}
\affiliation{
  \institution{Department of Computer Science and Technology, \\ Tsinghua University}
  \city{Beijing}
  \country{China}}

\author{Shwetak Patel}
\email{shwetak@cs.washington.edu}
\affiliation{
  \institution{Paul G. Allen School of Computer Science and Engineering, \\ University of Washington}
  \city{Seattle, WA}
  \country{USA}}

\author{Yuanchun Shi}
\email{shiyc@tsinghua.edu.cn}
\affiliation{
  \institution{Department of Computer Science and Technology, \\ Tsinghua University}
  \city{Beijing}
  \country{China}}
\affiliation{
  \institution{Qinghai University}
  \city{Xining, Qinghai}
  \country{China}}

\renewcommand{\shortauthors}{Yuntao Wang, et al.}

\begin{abstract}

A computer vision system using low-resolution image sensors can provide intelligent services (e.g., activity recognition) but preserve unnecessary visual privacy information from the hardware level. However, preserving visual privacy and enabling accurate machine recognition have adversarial needs on image resolution. Modeling the trade-off of privacy preservation and machine recognition performance can guide future privacy-preserving computer vision systems using low-resolution image sensors. In this paper, using the at-home activity of daily livings (ADLs) as the scenario, we first obtained the most important visual privacy features through a user survey. Then we quantified and analyzed the effects of image resolution on human and machine recognition performance in activity recognition and privacy awareness tasks. We also investigated how modern image super-resolution techniques influence these effects. Based on the results, we proposed a method for modeling the trade-off of privacy preservation and activity recognition on low-resolution images. 

\end{abstract}





\maketitle

\section{Introduction}
\label{sec:introduction}
The advances in technological engineering have enabled cameras to be increasingly ubiquitous.
Nowadays, many cameras can be manufactured at a low cost in a power-efficient manner, and with small sizes.
With the help of artificial intelligence, these cameras are enabled with automatic recognition abilities, providing smart services and applications publicly or privately~\cite{Applin-Amazon}. 
%. For instance, one prevalent application is to recognize people's daily behaviors and activities
However, in realistic scenarios, this brings up a major concern --- visual privacy exposure. 
We expect a vision-based system that can bring intelligent applications while preserving visual privacy. 

To achieve this purpose, researchers have explored many post-processing methods, which were often accomplished by decoupling the personally identifiable information (e.g., face)~\cite{visual-privacy-survey,Lang-privacy-design, Gross2009,Boult2005,Saini2014,Frome2009}.
However, these solutions are not sufficient to process all visual privacy cues~\cite{Orekondy-privacy-advisor, visual-privacy-survey,Ryoo2017, Ryoo-2016}. 

\begin{figure*}[!bp]
    \centering
    \begin{subfigure}{0.625\textwidth}
      \centering
      \includegraphics[width=1\textwidth]{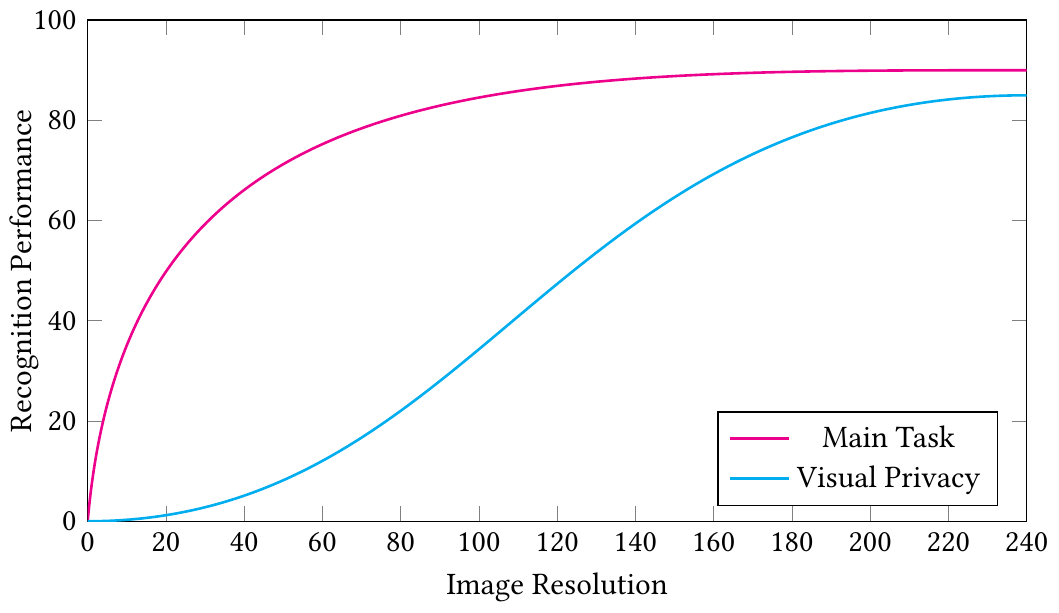}
    \end{subfigure}
    \hfill
    \begin{subfigure}{0.325\textwidth}
      \begin{subfigure}{0.475\textwidth}
        \includegraphics[height=1\textwidth]{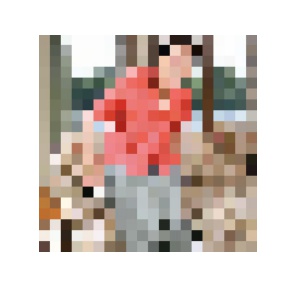}
        \caption{$20\times 20$}
      \end{subfigure}
      \begin{subfigure}{0.475\textwidth}
        \includegraphics[height=1\textwidth]{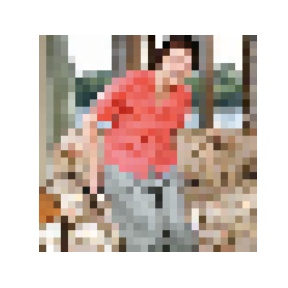}
        \caption{$30\times 30$}
      \end{subfigure}\\
      \begin{subfigure}{0.475\textwidth}
        \includegraphics[height=1\textwidth]{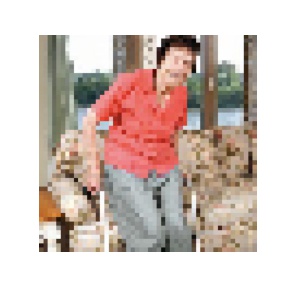}
        \caption{$50\times 50$}
      \end{subfigure}
      \begin{subfigure}{0.475\textwidth}
        \includegraphics[height=1\textwidth]{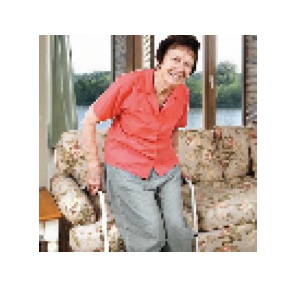}
        \caption{$100\times 100$}
      \end{subfigure}
    \end{subfigure}
    \caption{Demonstration of the effects of image resolution on the performance of the main vision-based recognition task and visual privacy awareness.}
    \label{fig:intro}
    \Description{Demonstration of the effects of image resolution on the performance of the main vision-based recognition task and visual privacy awareness. On the left figure, two curves indicate the trend of the main recognition task performance and visual privacy awareness performance over image resolution. On the right side, four images are displayed with resolutions of 20 * 20, 30 * 30, 50 * 50, and 100 * 100. In the image, an elder female is trying to stand up from a coach.}
\end{figure*}

As suggested by related works~\cite{Ryoo-2016, Ryoo2017, Ryoo2018, Xu-fully-coupled, Chen2016}, a fundamental solution toward the construction of a privacy-preserving vision-based system is to lower the image sensor's resolution from the hardware level. Thus, machines can achieve applicable performance in the main recognition task (e.g., activity recognition), while preserving visual privacy as much as possible.
Related works have proved that a low-resolution image (e.g., $16 \times 12$ pixels) possesses sufficient visual features for the main recognition task but not for visual privacy awareness. However, a high-resolution image can provide enough visual features for both of these two tasks. Thus, there is a trade-off regarding the effect of the image resolution on the main recognition task and visual privacy awareness as Figure~\ref{fig:intro} illustrates. Understanding and modeling such a trade-off will provide guidance for the privacy-preserving vision-based system with low-resolution visual sensors. 

In this paper, we focus on a smart home scenario where low-resolution image sensors automatically recognize activities of daily living (ADLs), such as feeding, entertainment, personal hygiene, intimacy, and functional mobility. ADLs recognition system can summarize activities and daily routines on which the ability of a person living independently is assessed; thus is widely used for health monitoring, especially for elderly care~\cite{Debes-ADL,Lawton-ADL}. 
% In a privacy-sensitive home environment, we expect the vision-based system accurately recognize the activity while preserving as much visual privacy as possible. 
% These two demands post adversarial needs over image resolution. 
% For instance, on images with high resolution (Figure~\ref{fig:intro}D), it is easy to recognize the activity as functional mobility. However, we can also easily identify the character's face and valuable properties (e.g., necklace). On images with ultra-low-resolution (Figure~\ref{fig:intro}A), it is hard for interpreters to recognize the privacy features but the activity either. Figure~\ref{fig:intro}B has a proper resolution, on which we can recognize the main activity but not the character's face or the property. Therefore, our goal is to propose a method to calculate an optimal resolution range for the visual privacy preserving ADLs recognition application. 
In realistic home environments, the data captured by an image sensor may be single-frame pictures~\cite{Ryoo2018, Chen2016} or multi-frame videos~\cite{Gao_MMTSA, Miyazaki2015}. We regard both of them as \textit{images} to model the trade-off between privacy preservation and activity recognition.

We considered such a trade-off as an optimization problem over image resolution. We conducted an online user survey with 115 participants to obtain the most important visual privacy features including nudity, identifiable face, valuable property, and relationship. 
In this paper, we regarded both the human and the machine as recognizers. 
Thus, we explored the effect of image resolution on both human and machines' ability in activity recognition and visual privacy awareness on the PA-HMDB51 dataset, which consists of over 500 videos from realistic environments~\cite{wu2019framework}. Specifically, we conducted a user study with 240 participants to investigate the effect of image resolution on human recognition performance. We evaluated the machine's performance on ADLs and visual privacy recognition tasks with cutting-edge machine learning approaches. Finally, we built a modeling method for calculating the trade-off of visual privacy preserving ADLs recognition using low-resolution images. We envision that our method can inspire other vision-based systems that require balancing privacy awareness and machine recognition performance.
Overall, the contributions of our paper are two-fold.

1) Using the at-home ADLs recognition as a scenario, we proposed a pipeline to investigate the effect of the image resolution on both human and machine performance on the main activity recognition task and visual privacy awareness. 

2) We presented a model for calculating the trade-off of visual privacy preserving activity recognition using low-resolution images. Using the proposed model, we can calculate an optimal resolution range of the image sensor for privacy-preserving activity recognition applications.

\section{Related work}
\label{sec:backgr-relat-work}
We describe the related work in this section, including visual privacy features and taxonomy, privacy-preserving machine recognition, and balancing the trade-off between privacy preservation and machine recognition.

\subsection{Visual Privacy Features and Taxonomy}
Privacy is described as "the right to select what personal information about me is known to what people" \cite{Westin-Privacy-and-freedom}. 
Pictures or videos convey a broad spectrum of privacy information, namely visual privacy. 
While legal and government entities legislated laws and policies on privacy protection \cite{privacy-1974, wiki-privacy-law}, their guidance leaves room for intruding visual privacy. 
Recently, researchers have explored the visual privacy exposure degree, visual privacy taxonomies/features, visual privacy importance, and visual privacy risk assessment using social media image databases~\cite{Li-Human-Perception, Orekondy_2018_CVPR, Orekondy-privacy-advisor}.
Orekondy et al. summarized 68 kinds of visual privacy features on social media images and then explored the feasibility of evaluating visual privacy exposure degree through machine learning approaches~\cite{Orekondy-privacy-advisor, Orekondy_2018_CVPR}. Li et al. summarized 7 categories, including 22 visual privacy features by crowdsourcing users' descriptions in their photo album~\cite{Li-Human-Perception}.
% Gurari et al. proposed 23 visual privacy features by summarizing images on the crowdsourcing platform for visually impaired people~\cite{Gurari_2019_CVPR}. 
These researches provide fundamental guidelines on taxonomy and the importance of visual privacy, which inspired us to design our user survey to explore the perceived importance of visual privacy in a home environment under varying image resolutions.  

\subsection{Privacy-Preserving Machine Recognition}

% We locate our work within the field of daily activity recognition in a home environment, specifically, activities of daily living (ADLs). ADLs are commonly used for health monitoring~\cite{Debes-ADL,Lawton-ADL}, summarizing activities and daily routines on which the ability of a person to live independently is assessed. Monitoring ADLs automatically is one key element in ambient assisted living (AAL) technologies, especially for elders. There are six basic ADLs: bathing/showering, dressing, feeding, functional mobility, personal hygiene, and continence. However, the home environment is a private space that contains lots of visual privacy information. Therefore, there is a strong demand for visual privacy preserving activity recognition for AAL techniques.
A growing number of privacy preserving computation technologies have emerged in recent years, which share the common promise of preserving privacy while also obtaining the benefits of computational analysis~\cite{agrawal2021computation}. 
% Generally speaking, within the field of activities of daily living (ADLs) recognition, there are two major categories of existing solutions: (1) wearable-based solutions~\cite{Pirsiavash2019, Debes-ADL, Pirsiavash2012, Nguyen-Egocentric-review} and (2) ambient sensing-based solutions~\cite{Debes-ADL, Eldib2016, visual-privacy-survey, Luo2017, Daher2017, Mashiyama2015}. Wearable solutions mainly rely on motion sensors~\cite{Pirsiavash2019} or egocentric cameras~\cite{Debes-ADL, Pirsiavash2012, Nguyen-Egocentric-review} to recognize activities accurately. However, they require devices to be attached to the user's body and suffer from limited battery life. Ambient sensing solutions, on the other hand, install devices in the home environment, such as RGB cameras~\cite{Eldib2016,visual-privacy-survey}, passive infrared (PIR) motion sensors~\cite{Luo2017}, depth cameras~\cite{Daher2017}, or low-resolution thermal image sensors~\cite{Mashiyama2015}. Among all these sensors, the optical image sensor attracted researchers due to its advantages of cost efficiency and ubiquitousness~\cite{Eldib2016,visual-privacy-survey}. 
To preserve visual privacy, existing solutions mainly adopted post-processing techniques such as image blurring and encryption techniques for images containing visual privacy information, e.g., human faces~\cite{visual-privacy-survey, Ilia-Face-Off, Lang-privacy-design,Ryoo2017, Gross2009, Boult2005, Saini2014, Frome2009}. However, these solutions are insufficient to protect all privacy information, including readable addresses, phone numbers, etc.~\cite{visual-privacy-survey, Ryoo2017, Ryoo-2016}. 
% Currently, in dynamic scenarios, post-processing methods to preserve all visual privacy cues are nearly possible~\cite{Orekondy-privacy-advisor}. 

Recently, researchers proposed a fundamental solution for a privacy-preserving vision-based system --- to lower the image sensor's resolution from the hardware level~\cite{Miyazaki2015, Ryoo2017, Ryoo2018, Xu-fully-coupled, Chen2016, Xu-pose-low-res}. 
Specifically, Miyazaki et al. developed a technology that can accurately detect the flow of people on low-resolution videos in which the faces cannot be distinguished~\cite{Miyazaki2015}. 
Dai et al. simulated a privacy-protected smart room prototype and then studied the performance impact of the image resolution from a single pixel to 10 $\times$ 10 pixels~\cite{Dai-privacy-activity}. They evaluated that five 10 $\times$ 10 resolution cameras can achieve a fairly high accuracy of 89.6\% on recognizing 9 human poses. 
Ryoo et al. proposed the inverse super-resolution (ISR) method for activity recognition on ultra-low-resolution videos, which also achieved state-of-art recognition accuracy while preserving identifiable personal information~\cite{Ryoo2017, Ryoo2018}.
% Xu et al. published a fully-coupled two-stream spatio-temporal network for human behavior recognition using extremely low-resolution videos~\cite{Xu-fully-coupled}.
%There are four important of the framework: 1) a fully-coupled network trained with high-resolution images to learn cross-domain transformation between high and low-resolution feature spaces; 2) a 3D convolutional components that extract compact and efficient spatio-temporal features for short video units; 3) a Recurrent Neural Network (RNN) for long-range temporal motion information; 4) two network streams for detailed motion features between adjacent two frames.
%As for the latest research, there is an on-going study conducted by Department of Computer Science and School of Medicine in Stanford University[28], which pays special attention on the privacy intrusiveness and violation issue caused by healthcare visual surveillance in hospitals. To solve this problem, they are building a privately-trained DCSCN super-resolution model that can enhance the utility of downsampled low-resolution depth images. Hopefully, this framework can guarantee a high degree of privacy while retaining enough utilities to perform healthcare-related human behavior detection.

These solutions showed the feasibility of activity recognition on low-resolution images. However, they only assumed an image resolution threshold (e.g., 10 $\times$ 10~\cite{Dai-privacy-activity}) to be able to preserve visual privacy without evidence. 
% Further, they failed to answer the question on "what's a proper resolution range for visual privacy preserving activity recognition?"
% only evaluated the machine's recognition performance with a preset image resolution without proper guidance (e.g., 10 $\times$ 10~\cite{Dai-privacy-activity}).
Obviously, the lower the resolution is, the better the visual privacy can be preserved. However, a lower resolution will inevitably decrease the amount of information for activity interpretation. It remains unknown how to balance the two adversarial demands on image resolution for recognizing the activity and safeguarding visual privacy. Our work is to answer this question by proposing a mathematical trade-off model and a method to calculate the optimal resolution range.

% \subsection{Effect of Image Resolution on Visual Privacy Awareness}

\subsection{Balancing Privacy Preservation and Machine Recognition}
%To evaluate the effect of image resolution on privacy awareness, a fine gain \xueyang{fine gain or fine-grained?} model is highly demanded. 
Researchers have discussed the trade-off between privacy preservation and activity recognition by quantifying humans' perceptions of privacy features. 
% An essential step in evaluating privacy preservation effects is to estimate people's ability to recognize privacy features with different privacy-preserving technologies. 
Some existing works have explored the effect of image resolution on human ability in facial recognition~\cite{Harmon-Masking, Yip-face-recognition}. Harmon and Julesz found that humans are good at facial recognition even when the portrait's resolution is down to 16 $\times$ 16 pixels~\cite{Harmon-Masking}. Yip and Sinha found that humans can still recognize celebrities' faces on portraits with a resolution of merely 7 $\times$ 10 pixels~\cite{Yip-face-recognition}. Some researchers also explored the impact of blur or pixelize filters at various levels on visual privacy awareness and activity recognition in the context of common workplace activities~\cite{Boyle-effects-of-filtered-video} or crowdsourced behavioral video coding~\cite{Lasecki-trade-offs}. They concluded the feasibility of achieving activity awareness while preserving visual privacy when tested on human eyes.

Taking human or the machine recognition performances into account, some researchers tried to understand how to balance privacy preservation and recognition performance. Alharbi et al. evaluated the effect of varying degrees of obfuscation on bystander privacy and visual confirmation utility~\cite{alharbi2019mask}. Hasan et al. studied the relative trade-offs between privacy (revealing and concealing selective attributes of objects) and utility (the visual aesthetics and user satisfaction of the image) of different image transforms~\cite{hasan2019experience}. Wu et al. formulated a novel adversarial training framework to learn anonymization transform for input videos such that the trade-off between target utility task performance and the associated privacy budgets is explicitly optimized on the anonymized videos~\cite{wu2019framework}.

However, these existing works have three limitations. First, They only tested humans' interpretation ability. An intelligent application relies on the machine for the main recognition task rather than the human. A more comprehensive study is highly demanded to explore how resolution affects both the human and the machine's recognition performance. Second, they mainly regarded the character's face as a privacy feature, which is insufficient to quantify a fine-grained model for privacy-preserving applications. Third, applying post-processing filters to high-resolution images differs from lowering the image sensor's resolution, which can preserve the visual privacy information from the hardware level with fewer on-device computing resources required.

Our work fills the gap mentioned above. We targeted enabling visual privacy-preserving machine recognition applications on low-resolution image sensors. We modeled the effects of image resolution on both the human and machine's ability in activity recognition and visual privacy awareness. Further, we proposed a quantitative survey method to model the importance of comprehensive visual privacy features.

%Closest to our work, some researchers explored the impact of blur or pixelize filter at various levels on both visual privacy awareness and activity recognition in the context of common workplace activities~\cite{Boyle-effects-of-filtered-video} or crowdsourced behavioral video coding~\cite{Lasecki-trade-offs}. However, they only consider face identification as the visual privacy, which is not sufficient to quantify a fine gain model for calculating the optimal resolution for visual privacy preserving applications. Besides, they only tested the performance of human recognition performance on activities. However, an intelligent application relies on the machine for the main recognition task rather than the human. 

\section{Problem Definition and Implementation Pipeline}
%The trade-off between image utility and privacy preserving has been widely discussed in the field of both computer vision and human-computer interaction. The general idea of various existing models is to preserve the users' privacy as much as possible without losing too much of the image utility. Here, we will define this trade-off problem in a more quantitative way.
This section offers the mathematical definition of privacy-preserving machine recognition using low-resolution images.

\subsection{Problem Definition}
Assume $\mathcal{X}$ to be the raw image set in a realistic environment that could be captured by the image sensor, for example, single-frame pictures or multi-frame videos. 
$f_r(\mathcal{X})$ represents the captured image set from the image sensor at a resolution of $r\times r$.
% $f_r(\cdot)$ denotes the function that re-samples the image set to a targeted image resolution at $r \times r$. 
% Thus, $f_r(\mathcal{X})$ represents that all images in the $\mathcal{X}$ set have the same resolution at $r \times r$.
Assume $T$ to be the main recognition task associated with $\mathcal{X}$, in this paper, ADLs recognition. $P$ is the visual privacy awareness task associated with $\mathcal{X}$. There are three main components in our model.
\begin{itemize}
    \item \textbf{Recognition Function}. We define the recognition function of the main recognition task $T$ as $f_T(\cdot)$ and the privacy detection function designed for the privacy feature $P$ as $f_P(\cdot)$. Both $f_T(\cdot)$ and $f_P(\cdot)$ can generate the recognition results given the captured image set $f_r(\mathcal{X})$. 
    % Our model will explore both human and machines' recognition performance on the image set. 
    To give an example of machine recognition, $f_T(\cdot)$ and $f_P(\cdot)$ can be computer vision models such as artificial neural networks.
    \item \textbf{Evaluation Function}. We define the evaluation function $L_T(\cdot)$ and $L_P(\cdot)$ which take both the outcome of the recognition function $f_T(f_r(\mathcal{X}))$ and $f_P(f_r(\mathcal{X}))$ as input and evaluate the performance of the recognizers according to the ground truth labels $g_T(\mathcal{X})$ and $g_P(\mathcal{X})$.
    \item \textbf{Importance Weights}. Considering the variety of privacy features contained in the captured image set, there may be differences in humans' perceived importance of different privacy features. For a given type of privacy feature $P_i$ in the privacy feature set $\mathcal{P}$, we define a weight coefficient $\omega_i$ to denote humans' perceived importance of $P_i$.
\end{itemize}

% \subsection{Optimization Objective}
% Empirically, when we set the resolution of the image sensor to a extremely low value, the performance of the privacy recognizer will drop significantly while the performance of the activity recognizer will not be affected too much. 
% Such an observation is one of the key assumptions of realizing the trade-off between action recognition and privacy preserving, which has been illustrated intuitively in Figure~\ref{fig:intro}. 
% To put it in a mathematical way, under the resolution of $r_1$ and $r_2$ where $r_1 \ll r_2$, then we define
% \begin{align*}
% \Delta L_T&=\left\vert L_T\left(f_T(f_{r_2}(\mathcal{X})), g_T(\mathcal{X})\right) - L_T\left(f_T(f_{r_1}(\mathcal{X})), g_T(\mathcal{X})\right)\right\vert\\
% \Delta L_P&= \left\vert L_P\left(f_P(f_{r_2}(\mathcal{X})), g_P(\mathcal{X})\right) - L_P\left(f_P(f_{r_1}(\mathcal{X})), g_P(\mathcal{X})\right)\right\vert 
% \end{align*}
% Our assumption is that the camera system can obtaining little detailed visual information while still capture as much necessary non-private information as possible under $r_1$ to maintain the performance of target image recognition task under $r_2$. In other words,
% \begin{align}
%     \Delta L_T \ll \Delta L_P
%     \label{eq:finding}
% \end{align}

\begin{figure*}[!bp]
    \centering
    \includegraphics[width=0.8\textwidth]{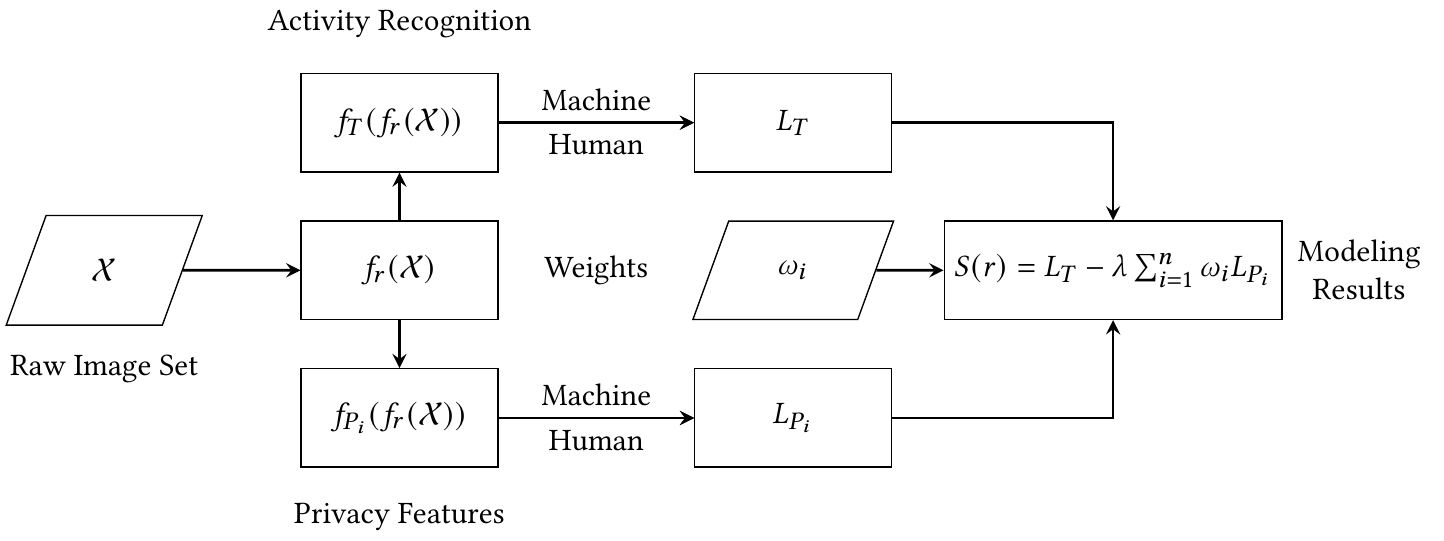}
    \caption{Framework for modeling the trade-off between privacy preservation and activity recognition.}
    \label{fig:framework}
    \Description{A flowchart introducing the framework for modeling the trade-off between privacy preservation and activity recognition. The raw image set was captured at the same resolution $r\times r$. Through a user study, we obtained users' perceived importance of different privacy features. For privacy feature and activity recognition tasks, we evaluate both machines' and humans' performances. Based on these studies, we then calculate the modeling results of the trade-off between privacy preservation and activity recognition.}
  \end{figure*}

% It is challenging to define such a trade-off that considers both activity recognition and privacy preservation. 
To optimize the trade-off between privacy preservation and activity recognition empirically, many prior works in computer vision have focused on finding suitable measurement metrics and objective functions mathematically~\cite{jihun2016minimax, jure2017closed, raval2017adversarial, wu2018pilot, wu2019framework}. However, most of these aforementioned works ignored humans' mental evaluation and recognition abilities of privacy features, and were thus insufficient. Based on prior works, we regard our research problem as mathematically optimizing the objective function $S(r)$ shown in Equation~\ref{eq:problem}, to reveal the trade-off between privacy preservation and machine recognition with both machine and human factors taken into consideration. 
\begin{equation}
    S(r)=L_T\left(f_T(f_r(\mathcal{X})), g_T(\mathcal{X})\right) - \lambda \sum_{i=1}^{n} \omega_i L_{P_i}\left(f_{P_i}(f_r(\mathcal{X})), g_{P_i}(\mathcal{X})\right)
    \label{eq:problem}
\end{equation}
Here, $\lambda > 0$ is a scaling factor representing the sensitivity ratio of visual privacy preservation over activity recognition performance.
The goal of formulating our research problem in the form of Equation~\ref{eq:problem} is to find out optimal resolution ranges where (1) cameras are limited from obtaining detailed visual information to preserve as much privacy information as possible and (2) cameras are able to capture as much detailed non-private information as possible to improve recognition performance.
 
\subsection{Implementation Pipeline}
The implementation pipeline of solving the optimization problem in Equation~\ref{eq:problem} has been depicted in Figure~\ref{fig:framework}.
% This section summarizes our core logic for solving this problem and describe the relationship between the different sections below.
In this paper, we choose the activities of daily living (ADLs) recognition task in a home environment as our target task $T$. 
% Nevertheless, it is still unclear which privacy feature $P$ we need to preserve preferentially given such a target task $T$. 
First of all, we conducted a user study to obtain humans' perceived importance of various privacy features ($\omega$ in our formulation), with  
% Also, it is crucial to know how human valuation of privacy features changes under different resolutions. 
the main results presented in section~\ref{sec:study1}. To evaluate our model in realistic environments, we utilized the publicly available video dataset PA-HMDB51 which is described in section~\ref{sec:dataset}.

The critical step of the whole implementation procedure is to model the recognition abilities of humans and machines under different resolutions. In order to preserve privacy comprehensively, we consider the recognition performance of both the human and the machine for each specific privacy feature $P$ to get an estimation of the evaluation results ($L_P$ in our formulation). The processing logic is similar for the activity recognition task to obtain the evaluation results ($L_T$ on our formulation). 
% Mathematically speaking, the recognition function in Equation~\ref{eq:problem} should be rewritten as
% \begin{align}
%     \widehat{f}_T&={\arg\max}_{f_T\in \mathcal{F}_T} L_T\left(f_T(f_r(\mathcal{X})), g_T(\mathcal{X})\right)\\
%     \widehat{f}_P&={\arg\max}_{f_P\in \mathcal{F}_P} L_P\left(f_P(f_r(\mathcal{X})), g_P(\mathcal{X})\right)
% \end{align}
% Take these results into Equation~\ref{eq:problem}, we can simplify our objective function as
% \begin{equation}
%     \widehat{S}(r) =L_T\left(\widehat{f}_T(f_r(\mathcal{X})), g_T(\mathcal{X})\right) - \lambda \sum_{i=1}^{n} \omega_i L_P^i\left(\widehat{f}_{P^i}(f_r(\mathcal{X})), g_P(\mathcal{X})\right)
%     \label{eq:implement}
% \end{equation}
We conducted a user study to model human recognition performance on those tasks in section~\ref{sec:study2}. Then, we utilized state-of-art computer vision models to finish those recognition tasks under different resolutions in section~\ref{sec:study3}. 
% However, with the development of super-resolution techniques, it is doubtful whether our estimation is robust against these state-of-art techniques. To handle this problem, we conducted additional surveys to check the robustness of our modeling results in section~\ref{sec:super}. 
Also, we provided additional analysis in section~\ref{sec:super} to check whether our modeling results are robust against currently state-of-art super-resolution techniques.
In the end, we proposed the calculating procedure of our objective function in section~\ref{sec:discussion} to model the trade-off between privacy preservation and activity recognition.

\section{Quantifying the Importance of Visual Privacy Features}
\label{sec:study1}
Our first user study aims to understand what visual privacy features users value the most and quantify the importance of those visual privacy features, thus simplifying the to-be-built model (Equation~\ref{eq:problem}). Inspired by related works~\cite{Orekondy-privacy-advisor,categories-of-privacy,Li-Human-Perception}, we obtained 25 visual privacy features that exist in a home environment. Here we divide them into 5 categories as below. 

\begin{itemize}
    \item \textbf{Biometric Identification}: identifiable face, gender, skin color, age group, weight group, hair color, eye color, and height group.
    \item \textbf{Personal Marker/Information}: nudity, home address, number/code, medical treatment, physical disability, handwriting, birthday, clothing, and tattoo.
    \item \textbf{Ethnicity}: religion, race, and nationality.
    \item \textbf{Society}: relationship, employment and pet.
    \item \textbf{Safety}: valuable property and living schedule.
\end{itemize}

\subsection{User Survey on Importance of Visual Privacy Features} 
We recruited 125 participants (66 females, 59 males) from MTurk. They had an average age of 32.7 (s.d. = 14.7). The whole survey lasted around 15 minutes. Each participant who passed the attention check received a 6 USD Amazon gift card. 

In the user survey hosted by Qualtrics~\footnote{https://www.qualtrics.com/}, we first introduced the smart home scenario where cameras are installed for ADLs recognition. Then we asked the participants to assume that they were living in the demonstrated house/apartment. 
% \input{tikz/rank_interface.tex}
% The user interface is illustrated in Figure~\ref{fig:rank_interface}. 
% On the one hand, each participant was required to finish the question of rating the importance of 5 privacy categories using 5-point Likert scale. The participant could click the options from the most important to the least important one by one with 1 for the most important and 5 for the least important. This is to explore how user value each category of visual privacy shown in the left column of Table~\ref{tab:privacy_category}. 
Then we evaluated the importance of each privacy feature with or without low-resolution to find out what privacy features users value the most and explore the effect of low-resolution on users' perceived importance of privacy features. 
For the high resolution test, we showed participants five high-resolution ($300 \times 300$) images. Each image captured one of the five basic daily activities: functional mobility, feeding, intimacy, entertainment, and personal hygiene. We did not control the participants' backgrounds regarding their culture, age, gender, and technical knowledge. In the instruction, we explicitly stated the scenario of visual privacy leakage as their similar pictures were posted on the Internet and thus can be accessed by everyone.
%\changed{We pointed to them explicitly that the images they saw in the questionnaire which contained certain types of privacy information could have been leaked to the public and possibly go everywhere on the Internet. Despite the difference in technical background of the participants which may have an impact on their perception of privacy as suggested by~\cite{kang2015everywhere}, we have tried out best to ensure that the way and content of privacy leakage understood by the participants are consistent in our study.} 
For the low-resolution test, we just showed participants the same five images in low-resolution ($50 \times 50$).
Under each resolution, the participant was asked how he/she values the importance of the different visual privacy information listed in the questionnaire. Then, the participants were required to rate the importance of each visual privacy feature using a 100-point slider where 0 stands for not important at all and 100 stands for extremely important. The score of each privacy feature shown on the slider updates along with the participant's choice. 

We designed two attention check questions under each condition. Each attention check question requires the participant to slide to a certain score that was generated randomly before each survey. All the questions were provided to the participant in random order. 
\begin{table*}[!ht] 
    \small
    \centering
    \caption{The statistic of the user rated importance scores of the 25 visual privacy features in 5 categories with and without the low-resolution conditions. $p < 0.05$ indicates significant difference between high and low resolution conditions. }
    \begin{tabular}
    {|M{0.2\textwidth}|M{0.17\textwidth}|M{0.07\textwidth}|M{0.07\textwidth}|M{0.07\textwidth}|M{0.07\textwidth}|c|}
    % {|c|c|c|c|c|c|c|}
    \hline
    
    \multirow{2}{*}{\textbf{Category}} & \multirow{2}{*}{\textbf{Feature}} & \multicolumn{2}{c|}{\textbf{High Resolution}} & \multicolumn{2}{c|}{\textbf{Low Resolution}} & \multirow{2}{*}{\textbf{Significance}} \\
    
    \cline{3-6} & & avg. & std. & avg. & std. & \\
    
    \hline
    \multirow{8}{*}{Biometric Identification} & Identifiable Face & 60.2 & 24.3 & \textbf{57.5} & 26.0 & $p = 0.13$\\
    \cline{2-7} & Gender & 43.5 & 29.2 & 43.4 & 29.4 & $p = 0.81$\\
    \cline{2-7} & Skin Color & 42.0 & 28.6 & 43.1 & 27.3 & $p = 0.94$\\
    \cline{2-7} & Age Group & 42.9 & 25.1 & 41.2 & 25.8 & $p = 0.35$\\
    \cline{2-7} & Weight Group & 43.9 & 27.2 & 40.9 & 27.2 & $p = 0.16$\\
    \cline{2-7} & Hair Color & 36.2 & 27.4 & 40.9 & 28.1 & $p = 0.05$\\
    \cline{2-7} & Eye Color & 40.4 & 28.9 & 40.3 & 28.4 & $p = 0.90$\\
    \cline{2-7} & Height Group & 37.3 & 25.8 & 40.0 & 27.7 & $p = 0.30$\\
      % \toprule
    \hline
    \multirow{9}{*}{\parbox{0.15\textwidth}{ \centering Personal Marker / Information}} & Nudity & 61.6 & 30.9 & \textbf{62.9} & 29.4 & $p = 0.71$\\
    \cline{2-7} & Home Address & 62.8 & 23.1 & 55.6 & 26.1 & $p = 0.01$\\
    \cline{2-7} & Number/code & 57.5 & 25.5 & 55.6 & 26.6 & $p = 0.79$\\
    \cline{2-7} & Medical Treatment & 60.4 & 23.2 & 51.7 & 25.9 & $p < 0.001$\\
    \cline{2-7} & Physical Disability & 52.1 & 25.1 & 49.4 & 26.0 & $p = 0.25$\\
    \cline{2-7} & Hand Writing & 52.6 & 26.4 & 44.9 & 27.7 & $p < 0.01$\\
    \cline{2-7} & Birthday & 54.2 & 26.8 & 44.7 & 28.5 & $p < 0.01$\\
    \cline{2-7} & Clothing & 40.5 & 27.9 & 41.5 & 27.5  & $p = 0.94$\\
    \cline{2-7} & Tattoo & 42.2 & 28.7 & 39.2 & 28.6 & $p = 0.34$\\
    \hline
    
    \multirow{3}{*}{Ethnicity} & Religion & 41.8 & 27.7 & 44.6 & 26.6 & $p = 0.29$\\ 
    \cline{2-7} & Race & 40.1 & 26.5 & 42.2 & 27.7 & $p = 0.64$\\
    \cline{2-7} & Nationality & 42.1 & 28.3 &  41.3 & 27.5 & $p = 0.46$\\
    \hline
    
    \multirow{3}{*}{Society} & Relationship & 60.3 & 24.8 & \textbf{52.9} & 25.7 & $p < 0.001$\\
    \cline{2-7} & Employment & 58.2 & 22.8 & 52.1 & 25.8 & $p = 0.05$\\
    \cline{2-7} & Pet & 37.3 &  24.4 & 39.1 & 27.8 & $p = 0.46$\\
    \hline
    
    \multirow{2}{*}{Safety} & Valuable Property & 64.0 & 25.0 & \textbf{59.6} & 26.1 & $p = 0.34$\\
    \cline{2-7} & Living Schedule & 59.3 & 24.4 & 59.1 & 26.3  & $p = 0.10$\\
    \hline
      % \bottomrule
    \end{tabular}
    \label{tab:privacy_importance}
  \end{table*}
\subsection{Result}
In total, we received 115 valid responses out of 120 total responses, in which respondents successfully completed the survey and passed all attention check questions. We utilized the Wilcoxon signed-rank test ($p < 0.05$) and Friedman test ($p < 0.05$) for statistical analysis since the rating scores are ordinal.

% Results reveal significance for the human perceived importance on different visual privacy categories ($\chi^2(4, N=115) = 37.6, p < 0.001$) using Friedman test. We used the Nemenyi test as a post-hoc test to illustrate the pairwise comparison between visual privacy categories. Results shown in table~\ref{tab:catergory_sig} indicate that people value bio-metric identification ($avg. = 3.53, s.d. = 1.33$) and personal marker/information ($avg. = 3.67, s.d. = 1.36$) more than other categories of visual privacy, including safety ($avg. = 4.00, s.d. = 1.34$), society ($avg. = 4.12, s.d. = 1.35$) or ethnicity ($avg. = 4.68, s.d. = 1.43$). Participants rated safety to be significantly more important than ethnicity. Further, society is significantly more important than ethnicity. \input{tables/category_sig.tex}

The analysis results are listed in Table~\ref{tab:privacy_importance}. We concluded with the following findings.

1. \textbf{Lowering the image sensor's resolution can significantly decrease users' concerns about visual privacy}. Table~\ref{tab:privacy_importance} shows the average and the standard deviation of the rating score of each visual privacy feature under two different resolution conditions. 
% Results show that users rated a significant lower score on the importance of visual privacy on low-resolution images ($Z=-4.02, p < 0.001$). 
On average, people rated visual privacy features with significantly lower importance scores ($Z=-4.02, p < 0.001$) under the low-resolution condition ($avg. = 45.1$) than the high-resolution condition ($avg. = 49.3$).

2. \textbf{Identifiable face, nudity, home address, number/code, medical treatment, relationship, employment, valuable property, and living schedule are considered to be more important than other visual privacy features.} Statistic analysis indicates that visual privacy features have significant effects on the human perceived important scores under either the high-resolution condition ($\chi^2(25, N=115) = 298.5, p < 0.001$) or low-resolution condition ($\chi^2(25, N=115) = 169.9, p < 0.001$). When we ran the pairwise statistical analysis using Wilcoxon signed-rank test among visual privacy features, we concluded with the following major results. On both high-resolution and low-resolution images, identifiable face, nudity, home address, number/code, medical treatment, relationship, employment, valuable property, and living schedule were considered the most important visual privacy features, since users rated them with significantly higher scores than other features ($p < 0.05$). Among these important privacy features, medical treatment and employment were considered less important ($p < 0.05$). 
%nudity has a significantly higher importance score ($avg. = 62.9, s.d. = 29.4$) than other visual features ($p < 0.05$) except identifiable face ($p = 0.25$), valuable property ($p = 0.24$) and living schedule ($p = 0.33$). 
%Participants rated identifiable face a higher importance score than other visual privacy features ($p < 0.05$) except home address ($p = 0.25$), nudity ($p = 0.25$), number/code ($p = 0.31$), valuable property ($p = 0.72$) and living schedule ($p = 0.74$). 
%Number/code is considered to be more important than other visual features ($p < 0.05$) except identifiable face ($p = 0.31$), relationship ($p = 0.15$), employment ($p = 0.20$), valuable property ($p = 0.24$) and living schedule ($p = 0.33$). 
%Home address is considered to be more important than other visual features ($p < 0.05$) except identifiable face ($p = 0.25$), number/code ($p = 0.77$), medical treatment ($p = 0.18$), physical disability ($p = 0.12$), relationship ($p = 0.18$), employment ($p = 0.26$), valuable property ($p = 0.16$). 
%Valuable property is considered to be more important than other visual features ($p < 0.05$) except identifiable face ($p = 0.72$), home address ($p = 0.16$), and living schedule ($p = 0.87$). 
%Living schedule is considered to be more important than other visual features ($p < 0.05$) except identifiable face ($p = 0.74$), nudity ($p = 0.33$), number/code ($p = 0.12$), and valuable property ($p = 0.87$). 

3. \textbf{Nudity, identifiable face, valuable property, and living schedule are the most important privacy features despite the image resolution.} When compared with the high-resolution condition, we observed significantly lower importance scores on features including home address ($p = 0.01$), medical treatment ($p < 0.001$), and relationship ($p < 0.01$) under the low-resolution condition. This finding is reasonable since these privacy features require high-resolution details to interpret. For instance, people were less concerned about the readable texts on low-resolution images. However, nudity, identifiable face, valuable property, and living schedule still lead to the most concerned visual privacy features in the low-resolution setting, with an average score above 57.

Instead of considering all the visual privacy features, we want to explore the most concerned ones that have the highest importance score and are potentially still vulnerable to low-resolution images. Therefore, we chose the most important visual privacy features in each category under the low-resolution condition with a minimum importance score threshold of 50.0. As a result, four visual privacy features including \textbf{\textit{nudity}}, \textbf{\textit{identifiable face}}, \textbf{\textit{valuable property}} and \textbf{\textit{relationship}} were chosen for later user studies and analysis. 
% We acknowledge that these four visual privacy features are limited to the scenario of daily activity recognition in the home environment. However, we envision that the research methods we introduced in this section can be applied to other scenarios. \xueyang{I didn't understand the last two sentences.}

\section{ADLs Dataset with Visual Privacy Features}
\label{sec:dataset}

This section describes the dataset we used to explore the effect of image resolution on humans' and machines' performance on activity recognition and visual privacy awareness tasks.

\subsection{Constructing the ADLs Dataset}

In order to evaluate the model in realistic environments, we used the publicly-available PA-HMDB51 dataset for privacy-preserving activity recognition~\cite{wu2019framework}. This dataset consists of about 355 minutes and 51 types of human activity videos collected from realistic environments with various visual privacy features annotated. 

In this paper, we mainly focus on activities of daily living (ADLs) in a smart home scenario. Therefore, three of our authors selected the qualified videos from the PA-HMDB51 dataset together with the following requirements. 1) The video represents a home environment. 2) All authors agreed that the main character conducted the same kind of activities. 3) All authors felt comfortable to publish the video online. For instance, due to the internet policy, we only chose men's or kids' topless videos in this study. Then, we divided the human activities in the PA-HMDB51 dataset into five basic kinds of activities of daily living (ADLs) including \textit{functional mobility}, \textit{feeding}, \textit{intimacy}, \textit{entertainment}, and \textit{personal hygiene}. Finally, we obtained 46, 30, 22, 37, and 16 minutes of videos for functional mobility, feeding, intimacy, entertainment, and personal hygiene, respectively.

We randomly split the PA-HMDB51 dataset into a training dataset, a validation dataset, and an evaluation dataset, which accounts for 90\%, 5\%, and 5\%, respectively. Considering the difference of the video duration in the PA-HMDB51 dataset, we divided all the videos into 2-second clips for later training and evaluation without affecting the judgment of the video content. Therefore, there are 226 clips of the videos in the evaluation dataset, with 69, 45, 33, 55, and 24 clips for functional mobility, feeding, intimacy, entertainment, and personal hygiene, respectively.

\subsection{Labeling the Privacy Features}

Based on the user study results presented in section~\ref{sec:study1}, we annotated each frame and each clip in our dataset with privacy features including \textit{nudity}, \textit{identifiable face}, \textit{valuable property}, and \textit{relationship}. 
Since privacy features may vary during the video clip, for example, even in the same video clip, the visibility of a person's face may be different, we provided both \textit{frame-level} and \textit{clip-level} labels of for each video in our dataset. First of all, we annotated all of the privacy attributes on each frame of different clips. Then, we annotated each clip according to the frames in the clip for later user studies and machine experiments. The detailed description of both frame-level and clip-level labels are listed below.
\begin{figure*}[!ht]
    \centering
    \begin{subfigure}{0.475\textwidth}
        \centering
        \includegraphics[width=1\textwidth]{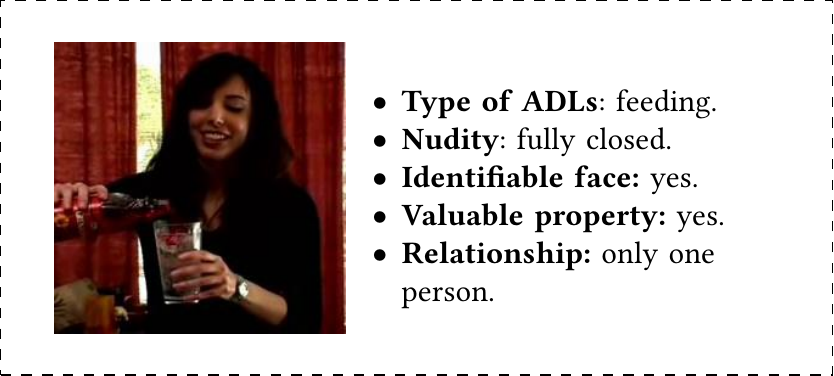}
    \end{subfigure}
    \begin{subfigure}{0.475\textwidth}
        \centering
        \includegraphics[width=1\textwidth]{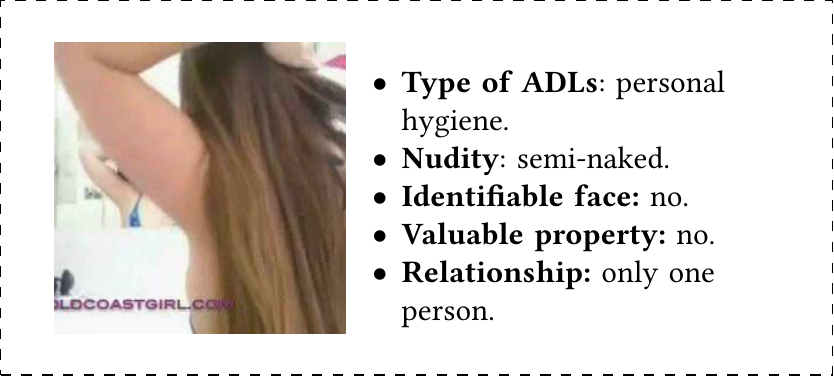}
    \end{subfigure} 
    \begin{subfigure}{0.475\textwidth}
        \centering
        \includegraphics[width=1\textwidth]{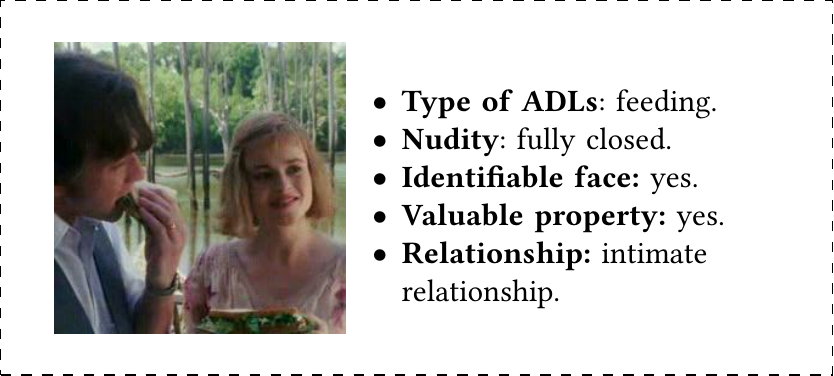}
    \end{subfigure}
    \begin{subfigure}{0.475\textwidth}
        \centering
        \includegraphics[width=1\textwidth]{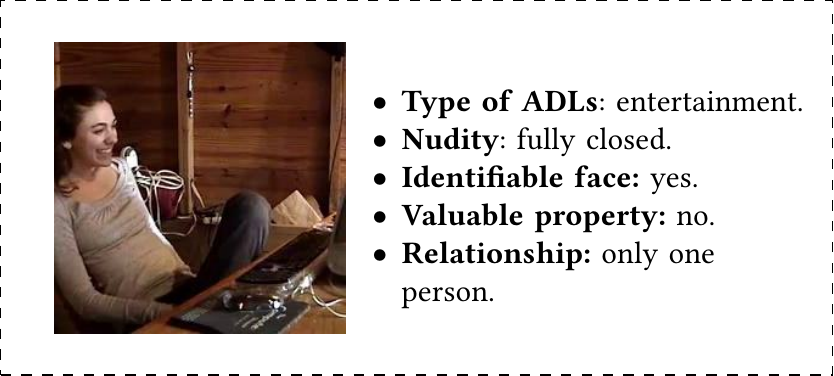}
    \end{subfigure} 
    \caption{Examples of the annotated frames in our dataset.}
    \label{fig:annotation_example}
    \Description{Examples of the annotated frames in our dataset. The four samples are shown in the upper left, upper right, lower left, and lower right of the figure, respectively. Each example is shown with a frame on the left, and labels of types of ADLs, nudity, identifiable face, valuable property, and relationship on the right.}
\end{figure*}
\begin{itemize}
    \item \textbf{Nudity}. The nudity label of each frame included three types that are \textit{naked or semi-naked (topless or bottomless)}, \textit{fully clothed}, and \textit{no person}. A clip is labeled as \textit{naked or semi-naked (topless or bottomless)} if at least one frame of the clip is labeled as \textit{naked or semi-naked (topless or bottomless)}. Otherwise, the clip is labeled as \textit{fully clothed} in a similar way. If every frame is labeled as \textit{no person}, we will finally label the clip as \textit{no person}.
    \item \textbf{Identifiable face}. If more than 70\% of a human face is visible, we consider the frame to contain an identifiable face. Therefore, each frame is labeled as \textit{yes}, \textit{no}, and \textit{no person}. A clip with more than one frame labeled as \textit{yes} is labeled as \textit{yes}, otherwise \textit{no}. A clip with every frame labeled as \textit{no person} is then labeled as \textit{no person}. 
    \item \textbf{Valuable property}. We only consider safe box, jewelry, watch, ring, and cash as valuable properties. Each frame is labeled as \textit{yes}, \textit{no}, and \textit{no person}. We label clips with at least one frame labeled \textit{yes} as \textit{yes}, otherwise \textit{no}. Clips with no person on any frame are labeled as \textit{no person}.
    \item \textbf{Relationship}. We consider the relationship of all the people presented in the video. There are four types of labels for each frame: \textit{intimate relationship}, \textit{non-intimate relationship}, \textit{only one person}, and \textit{no person}. A video clip is labeled as \textit{intimate relationship} if at least one frame of the clip is labeled as \textit{intimate relationship} and the frames labeled as \textit{intimate relationship} are no less than those labeled as \textit{non-intimate relationship}. Otherwise, a clip is labeled as \textit{non-intimate relationship} in a similar way. A clip with only one person presented is labeled as \textit{only one person} and labeled as \textit{no person} if there is no person existing in the clip.
\end{itemize}

Examples of the annotated frames in the dataset are demonstrated in Figure~\ref{fig:annotation_example}. Each frame was annotated by at least three of our authors and then cross-checked.
\section{Effect of Resolution on Human's Recognition Performance}
\label{sec:study2}
\begin{figure*}[!ht]
    \centering
    \includegraphics[width=1\textwidth]{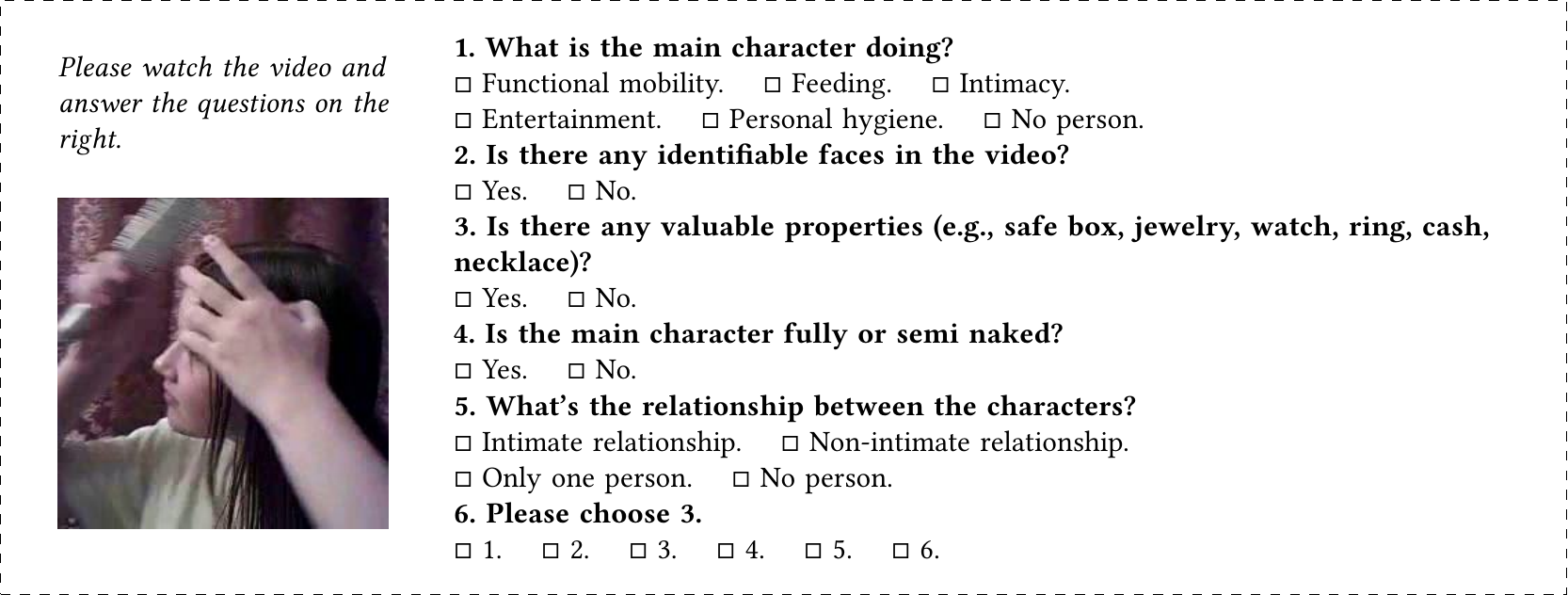}
  \caption{Example of the web-based user interface. Video clips of different resolutions is displayed on the left side. All tasks are listed on the right side of the web page.}
  \label{fig:us2_interface}
  \Description{Example of the web-based user interface. Video clips of different resolutions is displayed on the left side. Users are required to watch a video before answering the questions. All tasks including ADLs recognition, facial identification, nudity recognition, property detection, relationship classification, and attention checks are listed on the right side of the web page.}
\end{figure*}

After identifying the most important visual privacy features in the first study, we model the effect of image resolution on human performance in recognizing activities of daily living and visual privacy features. We describe the procedure and results in this section. 

\subsection{User Interface}
\label{sub:us2_user_interface_and_alg}

We developed a web-based user interface as shown in Figure~\ref{fig:us2_interface}. Each problem set in the test for the participants includes one ADLs recognition task and four privacy feature recognition tasks including face, nudity, valuable property, and relationship. The user interface also includes attention-check questions in each test. Responses with incorrect answers to the attention check questions were treated as invalid. A starting page, shown before the testing procedure, introduces the purpose of the user study and requires the participant's demographic information.

We sampled the image resolutions into seven values including $15 \times 15$, $20 \times 20$, $30 \times 30$, $50 \times 50$, $100 \times 100$, $160 \times 160$ and $240 \times 240$. We utilized a randomization strategy on the back-end server so that each participant could view 4 randomly chosen videos, with each video in a random resolution among these seven values. The same video did not appear twice to each participant. In addition, different clips from the same video did not appear to the same participant.

\subsection{Participant and Procedure}

We recruited 240 participants (105 females, 135 males) with an average age of 22.23 (s.d. = 5.25, ranging from 18 to 30). All participants were required to have healthy eye conditions without any historical disease (e.g., color blindness) and use their laptop or desktop web browser to finish the whole test. 
The starting page of the web-based user interface introduced the purpose of the study. 
Participants were required to fill in their demographic information, including gender, age, and historical eye diseases. Following were two practice tests using two $240 \times 240$ resolution example videos excluded from the evaluation dataset. Finally, each participant finished the 28 rounds of the test. The user study lasted around 10 minutes. Each participant was offered a 5 USD gift card for compensation. 

\subsection{Results and Findings}

In total, we obtained 6, 720 answer records, with 457 (6.80\%) invalid due to the failure of the attention check questions. We utilized One-way ANOVA for the statistic analysis ($p < 0.05$) with independent-samples t-test ($p < 0.05$) as post-hoc analysis. We present our major results and findings below.

\begin{figure}[!bp]
  \centering
  \begin{subfigure}{0.475\textwidth}
    \centering
    \includegraphics[width=1\textwidth]{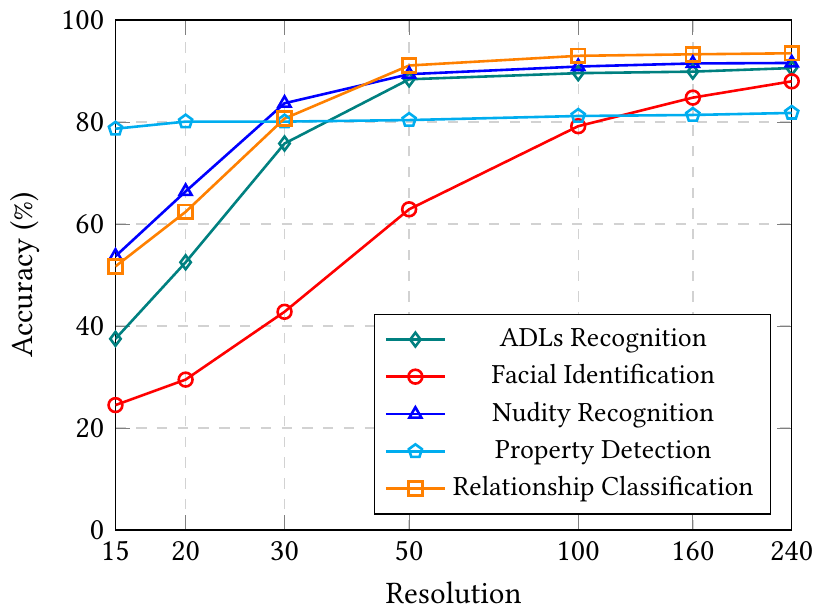}
  \end{subfigure}
  \caption{Humans' recognition performance on main activity and privacy feature recognition tasks.}
  \label{fig:us2_result}
  \Description{Humans' recognition performance on main activity and privacy feature recognition tasks. In the picture, the x-axis represents resolution and the y-axis represents accuracy. We plotted humans' recognition performance on tasks as curves including activity classification, facial identification, nudity recognition, property detection, and relationship classification.}
\end{figure}

\textbf{Low-resolution images are effective in preserving visual privacy but the effects are highly dependent on privacy features.} Figure~\ref{fig:us2_result} shows the effect of image resolution on human recognition performance of ADLs, face, valuable property, nudity, and relationship. We observed the significant effect of image resolution on all visual privacy recognition tasks ($p < 0.001$). 
Further, there is no significant difference between resolutions of $160 \times 160$ and $240\times 240$, indicating that resolutions above $160 \times 160$ pixels do not further contribute to visual privacy awareness statistically.
However, the effect of the image resolution is highly task-dependent. Statistical analysis indicates that the type of privacy features has significant effects on the perception performance ($F_{3,25048}=427.2$, $p < 0.001$). Specifically, pair-wise comparisons show that human eyes are more sensitive to nudity ($p < 0.001$) when the image resolution is below $50 \times 50$ pixels, followed by the relationship task. However, tasks including face identification and valuable property recognition require higher resolution images ($\geq 100 \times 100$ pixels) to achieve higher performance. For example, participants can only identify human faces with an accuracy of 79.2\% when the resolution is $100\times 100$ pixels. This is because both face identification and valuable property rely on detailed visual information. Therefore, a low-resolution image sensor can preserve but not fully protect visual privacy from the perspective of a human recognizer. 

\textbf{Lowering the image resolution has a significant negative impact on human recognition performance on ADLs.} Results show that there is a statistically significant effect of changed resolution on human ADLs recognition performance ($F_{6,6256}=278.0$, $p < 0.001$). With resolutions lower than $30\times 30$ pixels, human eyes can only recognize the ADLs with an accuracy below 75.8\%. When the image resolution increases to $50 \times 50$ pixels, participants can recognize the activity with a fair accuracy --- 88.4\%. However, participants are aware of some privacy features at the resolution of $50 \times 50$. For example, they can recognize the relationship and nudity with an accuracy of 91.1\% and 89.4\%, respectively.
\section{Effect of Resolution on Machine's Recognition Performance}
\label{sec:study3}

In this section, we explore the effect of image resolution on machine's recognition performance of ADLs and visual privacy features. We adopted the open-access cutting-edge deep learning methods as the machine recognizer. 
% We describe the machine learning approaches and the major results. 

\subsection{ADLs Recognition}
\label{sub:adl_machine}

\subsubsection{Training and Evaluation Dataset}
We applied data augmentation approaches to the training dataset in section~\ref{sec:dataset}, including horizontal flip, and Gaussian Noise, enlarging the dataset by four times. To fairly compare the recognition performance of the machine and the human, we utilized the same evaluation dataset in section~\ref{sec:study2}. 

\subsubsection{Training and Evaluation Procedure}
We utilized both convolutional neural networks and transformer-based models as our ADLs classifiers, including \textbf{ResNet50}~\cite{he2015deep}, \textbf{Efficient Net}~\cite{Mingxing2019EfficientNet} and \textbf{Vision Transformer (ViT)}~\cite{dosovitskiy2020vit}. All the models used here were pretrained with ImageNet dataset~\cite{imagenet} that output 1000 probabilistic values. In this experiment, we took every fame of the video clips in our dataset as the model input during our training, validating, and testing procedure. We first scaled the image of low resolution to $512\times 512$ pixels to standardize the input of the model. Then, we fine-tuned the pretrained network using the training dataset with a certain resolution ($r$) in which the images were all at the resolution of $r\times r$. To transfer the pretrained network model to our application, we added an additional five-node fully connected layer at the end of the network. We used sigmoid as the activation function. Once we finished the training procedure, we evaluated the fine-tuned model using the evaluation dataset under the same image resolution ($r$).
As we have described in section~\ref{sec:dataset}, we use the randomly chosen 5\% of the total dataset as the validation dataset in our implementation. In order to avoid the over-fitting problem, we used the early stopping method. In other words, we will stop our training procedure when the accuracy on the validation dataset does not rise anymore for 5 successive epochs. 

\subsubsection{Result}
Table~\ref{tab:us3_result} shows the effect of image resolution on machines' performance of the ADLs recognition task. Results indicate that \textbf{the machine outperforms the human regarding the ADLs recognition task on low-resolution images}. Vision Transformer can maintain an accuracy of 84.4\% even when the image resolution is as low as $20\times 20$. However, such a resolution is far from enough for humans to recognize ADLs at an ideal accuracy level. For resolutions above $100 \times 100$, both humans and machines can achieve a high accuracy above 90\%. Such results show the possibility of constructing a range of image resolutions to preserve visual privacy without bearing great loss in ADLs recognition simultaneously.

\begin{table*}[!ht] 
    \centering
    \caption{The ADLs recognition performance of ViT, ResNet50, and EfficientNet, compared with human.}
    \begin{tabular}{|M{0.2\textwidth}|M{0.15\textwidth}|M{0.15\textwidth}|M{0.15\textwidth}|M{0.15\textwidth}|}
    \hline
    \multirow{2}{*}{\textbf{Resolution}} & \multirow{2}{*}{\textbf{Human}} & \multicolumn{3}{c|}{\textbf{Machine}} \\ \cline{3-5} & & \multicolumn{1}{c|}{\textbf{ViT}} & \multicolumn{1}{c|}{\textbf{ResNet50}} & \textbf{EfficientNet} \\ \hline
    $15\times 15$  & 37.5\% & 81.0\% & 63.9\% & 52.9\% \\ \hline
    $20\times 20$  & 52.5\% & 84.4\% & 66.3\% & 63.5\% \\ \hline
    $30\times 30$  & 75.8\% & 89.8\% & 75.1\% & 68.0\% \\ \hline
    $50\times 50$  & 88.4\% & 90.7\% & 80.5\% & 74.6\% \\ \hline
    $100\times 100$ & 89.6\% & 92.2\% & 81.5\% & 75.1\% \\ \hline
    $160\times 160$ & 89.9\% & 93.2\% & 82.0\% & 80.0\% \\ \hline
    $240\times 240$ & 90.6\% & 94.6\% & 88.8\% & 83.9\% \\ \hline
    \end{tabular}
    \label{tab:us3_result}
\end{table*}
\begin{figure*}[htbp]
    \centering
    \begin{subfigure}{0.475\textwidth}
      \centering
      \includegraphics[width=1\textwidth]{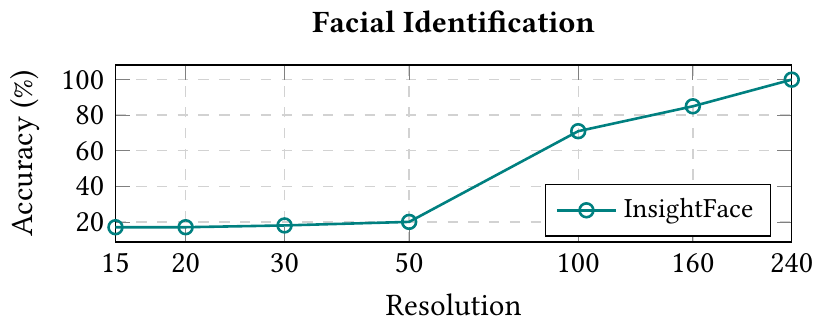}
    \end{subfigure}
    \begin{subfigure}{0.475\textwidth}
      \centering
      \includegraphics[width=1\textwidth]{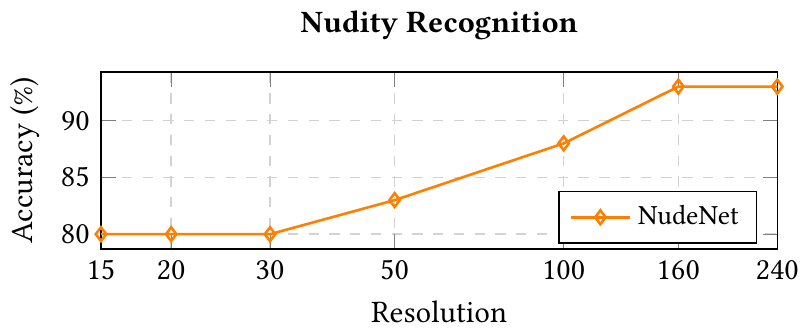}
    \end{subfigure}
    \begin{subfigure}{0.475\textwidth}
      \centering
      \includegraphics[width=1\textwidth]{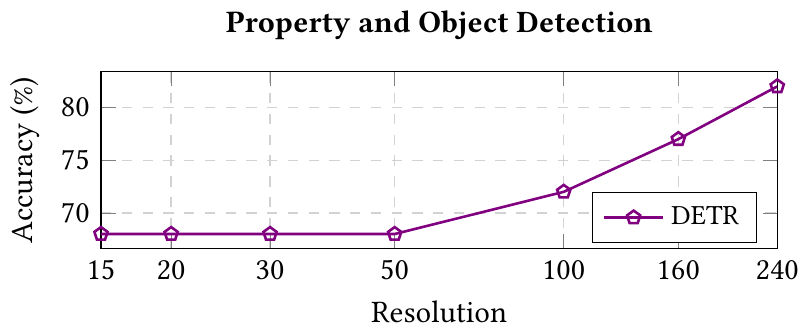}
    \end{subfigure}
    \begin{subfigure}{0.475\textwidth}
      \centering
      \includegraphics[width=1\textwidth]{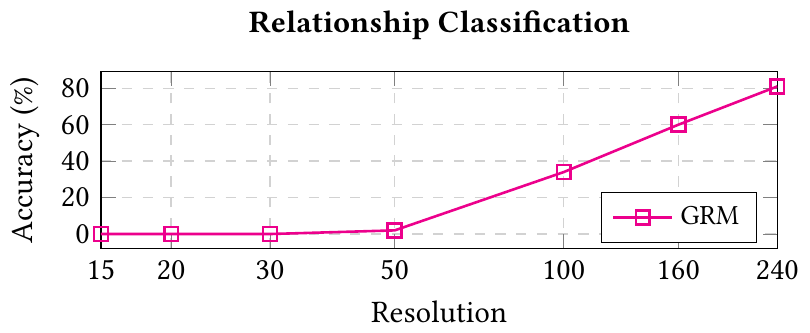}
    \end{subfigure}
    \caption{Machines' recognition performance of privacy features.}
    \label{fig:machine_result}
    \Description{Machines' recognition performance of privacy features. In the picture, the x-axis represents resolution and the y-axis represents accuracy. Shown in the upper left of the figure are the recognition results of InsightFace. Shown in the upper right of the figure are the recognition results of NudeNet. Shown in the lower left of the figure are the recognition results of DETR. Shown in the lower right of the figure are the recognition results of GRM.}
  \end{figure*}

\subsection{Privacy Features Recognition}

\subsubsection{Facial Identification}
\label{subsec:machine_face}
We adopted \textbf{InsightFace}~\cite{deng2018arcface} for facial identification by testing whether the model can recognize human faces in certain areas of the frames. We used the pretrained \textbf{ArcFace} model for facial identification provided by InsightFace. Also, we checked every frame of the video clips in this experiment. The result is shown in Figure~\ref{fig:machine_result} as the teal line. Results from the ArcFace model indicate that even the state-of-art models cannot detect any human faces below $50\times 50$ pixels. However, as the resolution increases from $100 \times 100$ to $240\times 240$ pixels, machine's facial identification performance significantly increases from 71.0\% to 100.0\%. Such results imply that identifiable faces can be preserved well against the machine attacker when the image resolution is below $50 \times 50$ pixels.

\subsubsection{Nudity Recognition}
\label{subsec:machine_nudity}
We adopted the pretrained \textbf{NudeNet}~\footnote{Software DOI: 10.5281/zenodo.3584720} for binary nudity recognition. This model was trained to detect nude parts of the human body in images. Here we utilized the classifier model to help us make a distinction between safe and unsafe images. We report the result of NudeNet as the orange line in Figure~\ref{fig:machine_result}. The precision and recall of NudeNet also reveal that it cannot identify any nude parts below the resolution of $30 \times 30$ pixels. Under the resolution of $100 \times 100$ pixels, NudeNet can recognize frames containing nude parts with an accuracy of 88.0\%. Therefore, we conclude that resolutions below $30 \times 30$ pixels can effectively preserve the nudity privacy feature.

\subsubsection{Property and Object Detection}
\label{subsec:machine_value}
We adopted \textbf{DETR}~\cite{Nicolas2020DETR} pretrained on the COCO dataset~\footnote{https://cocodataset.org/} for property and object detection. 
Considering the availability of pretrained object detection models, we used the detection performance of DETR on COCO objects as an estimation of machine's recognition performance on valuable properties.
We manually annotated the objects which belong to the COCO classes in each frame as ground truth. 
In our implementation, we first resized videos of different resolutions up to $240\times 240$ pixels. Then we kept bounding boxes with a confidence level above a pre-set threshold (e.g., 0.75) as a result of the model. 
To evaluate the model performance under different resolutions, we compared the objects detected by the model and the ground truth of each frame one by one to calculate the recognition accuracy. 
The purple line in Figure~\ref{fig:machine_result} shows the recognition accuracy of DETR. Results show that below the resolution of $50 \times 50$, DETR fails to detect any object. On images with a resolution of $100 \times 100$, DETR can achieve an accuracy of 72.0\%. Under the resolution of $160 \times 160$, large objects such as the main character can be detected precisely with an overall accuracy of 77.0\%. Under the resolution of $240\times 240$, the object detection results are more accurate, and small targets such as bottles and cups can be detected, too. Therefore, DETR can finally achieve an accuracy of 82.0\%.

\subsubsection{Relationship Classification}
\label{subsec:machine_relationship}

We adopted a pretrained cutting-edge social relationship classification model \textbf{GRM}~\cite{Wang2018Deep} on the evaluation dataset with different image resolutions. This model utilized a node message propagation mechanism and a graph attention mechanism to explore the interaction between the person pair of interest and contextual objects. The prerequisite to inferring the relationship between people is to obtain the context information using the object detection model. In our implementation, we resized the raw video of different resolutions to $240 \times 240$ pixels. Then, we annotated the bounding boxes and classes of different objects using DETR and labeled the bounding boxes of the person pair whose social relationship we wanted to examine. The model took every frame of the raw videos and objects list as input and generated the classification result as output.

The accuracy result we reported as the magenta line in Figure~\ref{fig:machine_result} describes the performance of the GRM model on the four-classes social relationship recognition task including \textit{intimate relationship}, \textit{non-intimate relationship}, \textit{no relationship}, and \textit{no person}. As is shown, the GRM model can detect nothing and will classify any input image as the \textit{no person} type under resolutions below $30 \times 30$ pixels. Our results here also proved that a low resolution below $30 \times 30$ is sufficient to preserve the privacy of social relationships against the cutting-edge machine recognition method. When the resolution is $100\times 100$ pixels, GRM can recognize social relationships in the video with an accuracy of 34.1\%. For resolutions of $160\times 160$ pixels and $240 \times 240$ pixels, GRM can achieve an accuracy of 60.9\% and 80.5\%, respectively.
\section{Justify the Influence of Image Super-Resolution}
\label{sec:super}

Image super-resolution techniques were proposed by researchers to reconstruct a high-resolution image from a low-resolution image~\cite{Wang2020, Liu_SpliteSR}. 
% Empirically, we expect super-resolution methods may have an impact on the recognition performance of humans and machines. 
In this section, we justice whether cutting-edge super-resolution techniques influence our results and findings regarding the effects of low resolution on activity recognition and privacy awareness through a user study.
%we conducted an additional user study to quantitatively investigate the effect of super-resolution. 
%the robustness of the recognition results of both humans and machines against several super-resolution techniques so as to provide a solid foundation for calculating the model in section~\ref{sec:discussion}. %So far, we have studied the recognition performance of humans and machines in section~\ref{sec:study2} and section~\ref{sec:study3}, respectively. %To our best knowledge, no existing work has thoroughly investigated the effect of super-resolution on humans' recognition performance. 

\subsection{User Study Procedure and Participant}
\begin{figure}[!ht]
    \centering
    \includegraphics[width=0.475\textwidth]{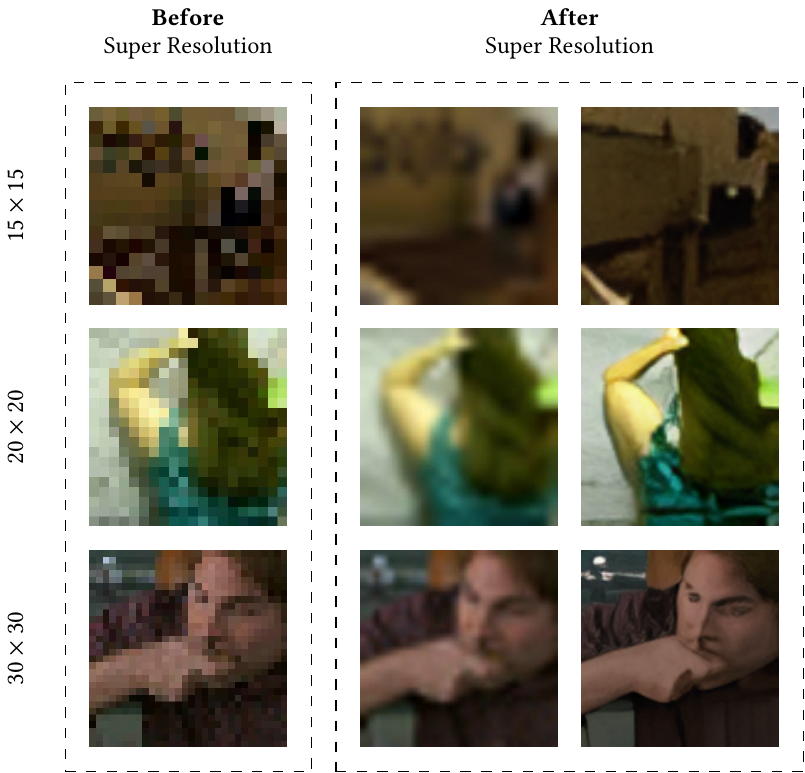}
    \caption{Examples of the effect of super resolution on videos of low resolutions including $15\times 15$, $20\times 20$, and $30\times 30$.}
    \label{fig:super_resolution}
    \Description{Examples of the effect of super resolution on videos of low resolutions including $15\times 15$, $20\times 20$, and $30\times 30$. Shown on the left side of the figure are frames not processed by super-resolution techniques. Shown on the right side of the figure are the same frames processed by super-resolution techniques.}
  \end{figure}

We adopted one of the cutting-edge image super-resolution methods SwinIR~\cite{Liang2021SwinIR} based on Transformer architectures as well as the traditional bicubic method to upscale the videos in our evaluation dataset by four times. Three examples of super-resolution processed videos are shown in Figure~\ref{fig:super_resolution}. 

We adopted a similar web-based interface as Figure~\ref{fig:us2_interface} shows except for changing the attention check question to addition and subtraction test. In this study, we first introduced the purpose and the procedure of our study. Then each participant took 8 trials with each trial having one test on the raw video and one test with videos after super-resolution. In each trial, we first presented each participant with a randomly-chosen raw video in the evaluation dataset and asked them to answer questions of ADLs and privacy features recognition as illustrated in Figure~\ref{fig:us2_interface}. The raw video's resolution was set to a random value among $15 \times 15$, $20 \times 20$, $30 \times 30$, $50 \times 50$, $100 \times 100$, $160 \times 160$ and $240\times 240$. Then, we presented them with the super-resolution videos together with raw videos simultaneously and asked them to answer the same questions. To avoid cross effects between videos under different resolutions, the same raw video did not appear twice to each participant. Further, we also ensured that the participants in this study were different from those who participated in the previous studies. 

We recruited 306 participants (123 females, 183 males) with an average age of 21.76 (s.d. = 4.56). The user study lasted around 10 minutes. Each participant was offered a 5 USD gift card for compensation. 

\subsection{Results and Findings}
In total, we received 4,896 test records with 273 (5.57\%) of them failed the attention check. Table~\ref{tab:super_resolution_adl} and Table~\ref{tab:super_resolution_privacy} show the comparison of participants' overall recognition accuracy with or without super-resolution. Results indicate that participants performed better on super-resolution videos than on raw videos. Statistical analysis suggests that when image resolution is below $20\times 20$ pixels, super-resolution techniques can significantly improve human recognition performance on both activity recognition and privacy recognition tasks. But it is worth noting that the improvement in recognition performance brought about by super-resolution technology is still less than that brought about by increasing the resolution itself. Such a finding reveals that super-resolution techniques do not provide enough additional information for humans to enhance their perception ability in both activity recognition and visual privacy awareness tasks. 

In terms of the impact of the super-resolution technique on the machine's recognition performance, researchers have proved that super-resolution can slightly facilitate vision-based recognition task such as activity recognition~\cite{demir2021tinyvirat, hou2021extreme},  object and text recognition~\cite{xi2020see, Liu_SpliteSR}. However, the influence of the super-resolution technique is very limited. The results are still significantly inferior to that with the original high-resolution images~\cite{Dai2015super}.

%scene recognition~\cite{Dai2015super}. Their experiment results illustrated that although super-resolution methods are helpful in general for other vision tasks when the resolution of input images are low, the performance with the super-resolved images are still significantly inferior to that with the original, high-resolution images.

In conclusion, the additional visual information introduced by the image super-resolution technique is insufficient to overcome the effect of resolution on the recognition performance of humans and machines. Therefore, we believe that the effects of image resolution on human (section~\ref{sec:study2}) and the machine's (section~\ref{sec:study3}) ADLs and visual privacy recognition performance are robust against image super-resolution techniques.

\begin{table}[htbp] 
    \centering
    \caption{The statistic of the overall accuracy on main activity recognition with or without super resolution conditions. $p < 0.05$ indicates a significant difference between with or without super resolution conditions. }
    \begin{tabular}
    {|M{0.1\textwidth}|M{0.0375\textwidth}|M{0.0375\textwidth}|M{0.0375\textwidth}|M{0.0375\textwidth}|M{0.1\textwidth}|}
    % {|c|c|c|c|c|c|}
    \hline
    \multirow{2}{*}{\textbf{Resolution}} & \multicolumn{2}{c|}{\textbf{Before}} & \multicolumn{2}{c|}{\textbf{After}} & \multirow{2}{*}{\textbf{Significance}} \\
    \cline{2-5} & avg. & std. & avg. & std. & \\ \hline
$15\times 15$                          & \multicolumn{1}{c|}{0.386} & 0.487 & \multicolumn{1}{c|}{0.452} & 0.498 & $p<0.001$                         \\ \hline
$20\times 20$                          & \multicolumn{1}{c|}{0.593} & 0.491 & \multicolumn{1}{c|}{0.706} & 0.456 & $p=0.002$                         \\ \hline
$30\times 30$                          & \multicolumn{1}{c|}{0.803} & 0.397 & \multicolumn{1}{c|}{0.845} & 0.362 & $p=0.149$                         \\ \hline
$50\times 50$                         & \multicolumn{1}{c|}{0.891} & 0.310 & \multicolumn{1}{c|}{0.893} & 0.308 & $p=0.932$                         \\ \hline
$100\times 100$                         & \multicolumn{1}{c|}{0.846} & 0.360 & \multicolumn{1}{c|}{0.898} & 0.302 & $p=0.046$                         \\ \hline
$160\times 160$                          & \multicolumn{1}{c|}{0.899} & 0.301 & \multicolumn{1}{c|}{0.908} & 0.289 & $p=0.701$                         \\ \hline
$240\times 240$                         & \multicolumn{1}{c|}{0.908} & 0.289 & \multicolumn{1}{c|}{0.927} & 0.260 & $p=0.386$                         \\ \hline

    \end{tabular}
    \label{tab:super_resolution_adl}
\end{table}
\begin{table}[htbp] 
    \centering
    \caption{The statistic of the overall accuracy on privacy features recognition with or without super resolution conditions. $p < 0.05$ indicates a significant difference between with or without super resolution conditions. }
    \begin{tabular}
    {|M{0.1\textwidth}|M{0.0375\textwidth}|M{0.0375\textwidth}|M{0.0375\textwidth}|M{0.0375\textwidth}|M{0.1\textwidth}|}
    % {|c|c|c|c|c|c|}
    \hline
    \multirow{2}{*}{\textbf{Resolution}} & \multicolumn{2}{c|}{\textbf{Before}} & \multicolumn{2}{c|}{\textbf{After}} & \multirow{2}{*}{\textbf{Significance}} \\
    \cline{2-5} & avg. & std. & avg. & std. & \\ \hline
    $15\times 15$ & \multicolumn{1}{c|}{0.558} & 0.497 & \multicolumn{1}{c|}{0.602} & 0.476 & $p<0.001$ \\ \hline
    $20\times 20$ & \multicolumn{1}{c|}{0.673} & 0.469 & \multicolumn{1}{c|}{0.736} & 0.440 & $p<0.001$ \\ \hline
    $30\times 30$ & \multicolumn{1}{c|}{0.793} & 0.404 & \multicolumn{1}{c|}{0.823} & 0.381 & $p=0.038$ \\ \hline
    $50\times 50$ & \multicolumn{1}{c|}{0.851} & 0.356 & \multicolumn{1}{c|}{0.866} & 0.340 & $p=0.276$ \\ \hline
    $100\times 100$ & \multicolumn{1}{c|}{0.895} & 0.305 & \multicolumn{1}{c|}{0.906} & 0.291 & $p=0.359$ \\ \hline
    $160\times 160$ & \multicolumn{1}{c|}{0.905} & 0.292 & \multicolumn{1}{c|}{0.913} & 0.280 & $p=0.488$ \\ \hline
    $240\times 240$ & \multicolumn{1}{c|}{0.921} & 0.268 & \multicolumn{1}{c|}{0.925} & 0.263 & $p=0.766$ \\ \hline
    \end{tabular}
    \label{tab:super_resolution_privacy}
  \end{table}
\section{Modeling the Trade-off of Privacy Preservation and Activity Recognition}
\label{sec:discussion}

In this paper, our goal is to present a method to model the trade-off between privacy preservation and machine recognition. We have obtained the estimation results of the main components in Equation~\ref{eq:problem}. In this section, we take all these results into consideration and explain how we can utilize them to model the trade-off between privacy preservation and machine recognition. Based on our modeling results, we further present how to apply our model to applications.

\subsection{Build the Model Using the Parameters from the Studies}

To summarize, we have investigated users' perceived importance of different privacy features under high or low image resolutions in section~\ref{sec:study1}. We chose users' rating of these privacy features under high-resolution image condition as the importance weight $\omega_i$ in the model, which was shown in Table~\ref{tab:privacy_importance}. 
Next, we examined both human's and the machine's recognition performance under varying resolutions in order to obtain an approximation of the evaluation function $L_T$ and $L_P$ in our formulation.
% Section~\ref{sec:dataset} described the dataset we used to explore the effect of image resolution on human's and the machine's performance. 
% In section~\ref{sec:study2}, we conducted a user study to understand human performance in recognizing main activities and privacy features. 
% In section~\ref{sec:study3}, we utilized open-access cutting-edge deep learning methods to explore the machine's recognition abilities on the same tasks. 
In realistic environments, intelligent applications may rely on either humans or machines to obtain private information from raw images. Therefore, we take both human and machine recognizers into consideration to preserve privacy features in a comprehensive way. For the main recognition task $T$, which is activity recognition in our implementation, the Vision Transformer outperforms all other models including humans even on extremely low-resolution videos from the dataset. Therefore, we choose the Vision Transformer as our final recognition function $f_T$ and the evaluation results of the Vision Transformer $L_T$ have been demonstrated in Table~\ref{tab:us3_result}. For each privacy feature $P_i$ including nudity, identifiable face, valuable property, and relationship, we found that humans are generally more effective recognizers compared with machines, especially on ultra-low-resolution videos from the dataset. Therefore, we consider humans as the final $f_{P_i}$ in our calculation. The evaluation results of each  $L_{P_i}$ we are going to use has been depicted in Figure~\ref{fig:us2_result}.

\subsection{Calculating the Objective Function}

\begin{figure*}[!ht]
    \centering
    \begin{subfigure}{0.325\textwidth}
        \centering
        \includegraphics[width=1\textwidth]{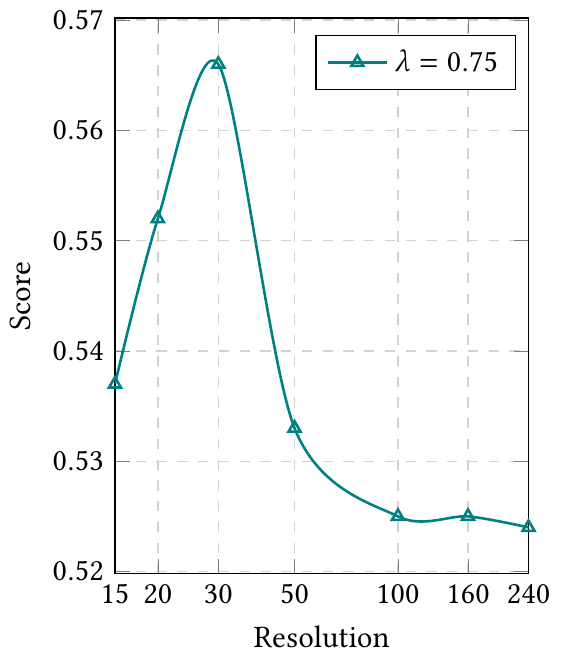}
    \end{subfigure}
    \begin{subfigure}{0.325\textwidth}
        \centering
        \includegraphics[width=1\textwidth]{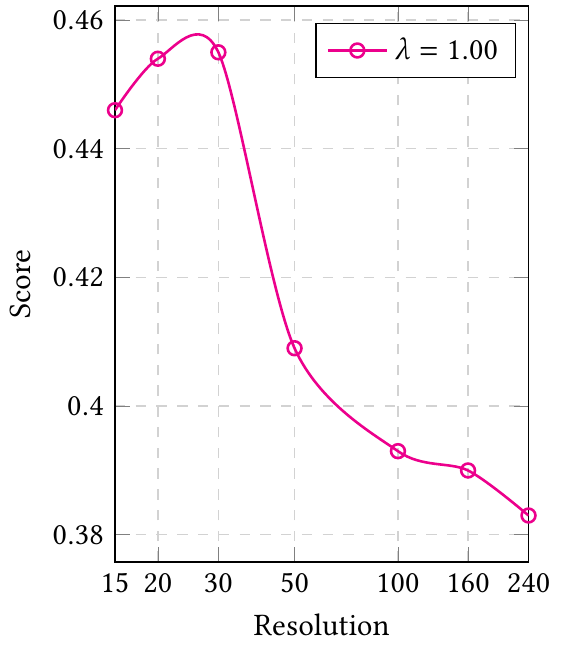}
    \end{subfigure}
    \begin{subfigure}{0.325\textwidth}
        \centering
        \includegraphics[width=1\textwidth]{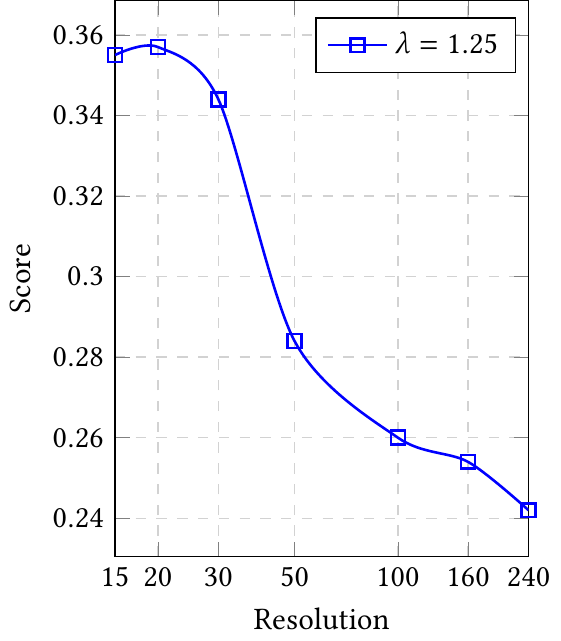}
    \end{subfigure}
    \caption{Depicting the objective function based on the results of both humans' and machines' recognition performance.}
    \label{fig:calculation_result}
    \Description{Depicting the objective function based on the results of both humans' and machines' recognition performance. Presented are three cases of the objective function where $\lambda$ equals 0.75, 1, and 1.25, respectively.}
\end{figure*}

Based on the results of $L_T$, $L_{P_i}$, and $\omega_i$ we have discussed above, we can calculate the objective function $S(r)$ in Equation~\ref{eq:problem} for each resolution we have sampled. Figure~\ref{fig:calculation_result} illustrates how the values of our objective function $S(r)$ change with resolutions $r$. The scaling factor $\lambda$ in our formulation indicates the sensitivity ratio of privacy preservation over activity recognition which can be flexibly adjusted according to the deployment environment or user experience. Here we have only shown the cases for three different lambda values, including $0.75$, $1.00$, and $1.25$. 

As is demonstrated in Figure~\ref{fig:calculation_result}, the value of the objective function $S(r)$ shows a trend of first increasing and then decreasing with the increase of resolution $r$. For the case where lambda is $1.00$, the objective function takes its maximum value at a resolution between $20\times 20$ and $30\times 30$, which indicates a proper resolution for balancing privacy preservation and activity recognition. Such an image resolution value can be easily extended to a certain image resolution range where the trade-off result is also acceptable. However, the objective function takes a low value when the resolution is too low (e.g., $15\times 15$) or too high (e.g., $240\times 240$). The reason behind this is also consistent with our expectations. When the image resolution is too low, although the privacy features can be better preserved, the machine's ADLs recognition performance is far from satisfactory. On the contrary, high image resolution may greatly increase the risk of privacy feature leakage except for improving ADLs recognition performance. 

Here we also noticed that as the scaling factor $\lambda$ increases, the maximum point of the objective function is also shifted to the left in Figure~\ref{fig:calculation_result}. Such a finding shows that a lower resolution of the image sensor is required if users are more concerned with privacy preservation compared with activity recognition performance. 

\subsection{Applying the Model and the Modeling Method to Applications}
In this section, we present how to apply our method and model to privacy-preserving machine recognition applications.

\subsubsection{Deployment to a Real Scenario Application}
When deploying a real scenario application based on our method, one can install an ultra-low-resolution (e.g., $20 \times 20$ pixels) image sensor with an edge computer running a machine learning method for ADLs recognition at home. To apply our framework for quantifying the trade-off between privacy preservation and activity recognition, one first needs to determine the sensitivity indicator $\lambda$ in Equation~\ref{eq:problem}, which is closely related to deployment environment and user experience. In our ADLs recognition example, the bathroom is a more visual privacy-sensitive location than the kitchen. Therefore, we would expect the image sensor in the bathroom having a lower resolution to preserve more visual privacy. 
Second, with the development of computer vision technologies, the performance of machine recognition on both activities and privacy features will exceed the current results stated in this paper. Future designers just need to fine-tune the results of the evaluation function $L_T$ and $L_P$ by selecting better recognizers $f_T$ and $f_P$ to consider the results of these technological advances.  
Third, one can leverage activities' probability distribution regarding the different environments in a home environment which may have an effect on the results of the evaluation function $L_T$ and $L_P$ in our formulation. For instance, personal and toilet hygiene is highly possible to happen in the restroom, while feeding is highly possible to occur in the kitchen. Future designers need to modify their training and evaluation data set according to the probability distribution of activities of daily living (ADLs) in different scenarios.

\subsubsection{Generalization to Other Applications}
For other computer vision based applications in a real scenario, we believe that our pipeline and method can be easily adapted. For instance, using an always-on low-resolution camera on AR glasses for activity recognition, or using a low-resolution smartphone camera for hand gesture recognition, etc. Even though different applications have their own usage scenarios with different visual privacy features, our method's key idea and basic framework can still be used efficiently. 
Although low-resolution image sensors can preserve visual privacy from the hardware level, deploying the hardware itself costs a large of human labor and money. Instead of purchasing the low-resolution image sensor, we can simply update the firmware to limit the camera's resolution, turning them to low-resolution image sensors. Further, we can attach an additional layer or lens on top of available commercial RGB cameras. For instance, we can add a piece of frosted glass, or a lens built for the Passive Infrared (PIR) motion sensor to the commodity cameras ~\footnote{https://en.wikipedia.org/wiki/Passive\_infrared\_sensor}. Most of these camera system parameter selection issues can be discussed and solved in a more generalized way of our methods.

\section{Limitations and Future Work}

Our work is targeted at modeling the trade-off between visual privacy and the core machine recognition task, e.g., activity recognition in our case. Our purpose is to inspire future work to explore more quantitative methods for privacy-preserving applications. Therefore, future designers can apply or adapt these models according to their applications to preserve users' privacy as much as possible. However, there do exist several limitations of our work and we describe them below.

\paragraph{Utilizing Multimodal Information}

We acknowledge that an image sensor deployed at home can only collect images at a fixed position, distance, and field of view after installation. 
Only with the single modality data captured by images sensors, both machine's and human's recognition performance can be easily affected by the aforementioned factors.
We also acknowledge that we didn't take multimodal data, for example, audio data into consideration. Prior works have proved the effectiveness of leveraging multimodal data in activity recognition. With multimodal information, we can alleviate existing algorithms' dependence on images, thus allowing for a lower resolution of image sensors.
We expect future research can investigate how the modeling results of the trade-off between privacy preservation and activity recognition can be changed by multimodal information.

\paragraph{Privacy Preserving Methods}

In this work, we only use pixelization filters as the privacy-preserving method for the main task. The advantages of using low-resolution images have already been discussed in prior works. Nevertheless, we have to admit that researchers have shown that low resolution alone does not provide enough privacy guarantees. McPherson et al. found that obfuscated images contain enough information correlated with the obfuscated content to enable accurate reconstruction of the latter~\cite{McPherson2016defeating}. Although we have compared the privacy recognition performance of state-of-art machine learning algorithms on low-resolution images, we believe that our evaluation results on low-resolution images leave much room for discussion. We expect future research can explore the effect of more privacy-preserving methods on the trade-off between privacy preservation and activity recognition.

\paragraph{User Survey on Importance of Visual Privacy Features}

We acknowledge that our user study in section~\ref{sec:study1} aims to assess users' perceived importance of visual privacy features. We didn't limit participants' culture, age, gender, or technical backgrounds. However, there are many other factors that may affect participants' perception of privacy. For example, researchers have found that users on Amazon Mechanical Turk, where our participants were from, tend to be more privacy conscious~\cite{kang2014mturk,ross2010demographics}, thus are not representative of the general population all over the world. 
It is also undeniable that the perception of privacy varies substantially across cultures, societies, times, and locations~\cite{albayaydh2022jordan, ahmed2017digital-privacy, ahmed2017sim, crabtree2017repacking, palen2003unpacking, sambasivan2018rich, kang2015everywhere}. Therefore, our estimation of the perceived importance ($\omega$ in our formulation) of privacy features obtained through our user studies is possibly not applicable to populations in different cultural contexts across the world. However, the framework proposed in this paper is meant to inspire future researchers to consider humans' assessments of the importance of different visual privacy features. We expect that there will be more independent works to explore the influence of other factors on humans' perception of privacy.

\section{Conclusion}
\label{sec:conclusion}
Using the at-home activity of daily livings (ADLs) as the scenario, this paper models the trade-off of visual privacy preservation and activity recognition over image resolution. To achieve this purpose, we first conducted a user survey to obtain the most important visual privacy features, including nudity, identifiable face, valuable property, and relationship. Then, using the PA-HMDB51 dataset, which contains videos from realistic environments, we quantified the effect of image resolution on the human's performance on ADLs recognition and visual privacy awareness tasks through a user study. We further modeled the impact of image resolution on the machine's ability to recognize ADLs and visual privacy features using cutting-edge machine learning methods. Finally, we proposed a method with adjustable parameters to model the trade-off of privacy-preserving ADLs recognition using low-resolution images. Using this method, we can calculate an optimal range of image resolution for visual privacy preserving ADLs recognition. We envision that our method can inspire other vision-based systems that require balancing privacy awareness and machine recognition performance.

\begin{acks}
This work is supported by the Natural Science Foundation of China (NSFC) under Grant No. 62002198 and No. 62132010, Tsinghua University Initiative Scientific Research Program, Beijing Key Lab of Networked Multimedia, Institute for Artificial Intelligence, Tsinghua University, and Beijing National Research Center for Information Science and Technology (BNRist).
\end{acks}

% Bibliography
\bibliographystyle{ACM-Reference-Format}
\bibliography{proceedings}


\end{document}